\documentclass[12pt]{iopart}
\usepackage{tikz}
\usepackage{units}
\usepackage{setstack}
\usepackage{nicefrac}
\usepackage[colorlinks]{hyperref} 
\usepackage{amstext}
\usepackage{cite}
\usepackage{iopams}  

\usepackage{subcaption}
\usepackage{multirow}

\usepackage{graphicx}
\begin{document}

\title[Huang \& Marian]{A Generalized Ising Model for studying Alloy Evolution under Irradiation and its use in Kinetic Monte Carlo Simulations}

\author{Chen-Hsi Huang$^1$ \& Jaime Marian$^2$}

\address{University of California Los Angeles}
\ead{\\
$^1$skyhuang@ucla.edu\\
$^2$jmarian@ucla.edu}
\vspace{10pt}

\date{\today}

\begin{abstract}
We derive an Ising Hamiltonian for kinetic simulations involving interstitial and vacancy defects in binary alloys. Our model, which we term `ABVI', incorporates solute transport by both interstitial defects and vacancies into a mathematically-consistent framework , and thus represents a generalization to the widely-used ABV model for alloy evolution simulations. The Hamiltonian captures the three possible interstitial configurations in a binary alloy: A-A, A-B, and B-B, which makes it particularly useful for irradiation damage simulations. All the constants of the Hamiltonian are expressed in terms of bond energies that can be computed using first-principles calculations.  
We implement our ABVI model in kinetic Monte Carlo simulations and perform a verification exercise by comparing our results to published irradiation damage simulations in simple binary systems with Frenkel pair defect production and several microstructural scenarios, with matching agreement found. 
\end{abstract}

\pacs{61.50.Ah, 02.70.Uu, 61.72.Bb, 64.75.Op, 61.80.Az}
%
%
\submitto{\JPCM}
%
%
%

\section{Introduction}
Stochastic modeling of microstructural evolution in substitutional binary alloys using Monte Carlo methods is a relatively mature field. In lattice kinetic Monte Carlo simulations,  alloy configurations are generated randomly, typically by direct atom exchange (the so-called `Kawasaki' dynamics) \cite{1966PRKawasaki, 2000Domb, 2000PASMaiAFratzl, 2003EELWeinkamer, 2004PTWeinkamer}, or by (local) vacancy-mediated solute transport \cite{2000PASMaiAFratzl, 2003EELWeinkamer, 2004PTWeinkamer, 1991JoSPYaldram, 1991AMeMYaldram, 1998PRBWeinkamer, 2002CMSSchmauder, 2007PRBSoisson, 2008JoNMVincenta, 2011PRBReina, 2012CMSWarczok}. The time scale is recovered by using physical jump frequencies that depend on the energies of the configuration before and after the exchange in such a way that detailed balancing holds. These energies are calculated using a suitable Hamiltonian function, which --in most cases-- depends only on the chemical nature of the species participating in an exchange, as well as on their separation distance. Such methods, aptly called `AB' or `ABV' --in reference to the atomic species involved--, generally express the Hamiltonian as a cluster expansion truncated to first or second nearest neighbor distances \cite{2000PASMaiAFratzl, 2003EELWeinkamer, 2004PTWeinkamer, 1991AMeMYaldram, 1998PRBWeinkamer, 2008JoNMVincenta, 2011PRBReina, 2010PRBLavrentiev}. The order of the cluster expansion is variable, although it is generally restricted by computational considerations to second order \cite{2000PASMaiAFratzl, 2003EELWeinkamer, 2004PTWeinkamer, 1991JoSPYaldram, 1998PRBWeinkamer, 2011PRBReina, 2012CMSWarczok, 2010PRBLavrentiev}. However, it is often advantageous to express the cluster expansion Hamiltonian in terms of an Ising model where the site occupancy variables reflect the nature of the different species involved. This is because of the extensive mathematical and computational infrastructure associated with the Ising system, which is one of the most widely studied, and whose behavior is best understood, models in computational physics \cite{2000PASMaiAFratzl, 2003EELWeinkamer, 2004PTWeinkamer, 1998PRBWeinkamer, 2010PRBLavrentiev, 1992STATSYeomans, 2002MONTEBinder, 1967RMPBrush}. 

ABV models are also of interest in irradiated materials, to study non-equilibrium phenomena such as radiation enhanced diffusion and segregation, and indeed have been applied on numerous times in irradiation damage scenarios \cite{1996AMSoisson, 2001PRBEnrique, 2003JoAPEnrique, 2008JoNMVincenta, 2013PRBShu}. 
However, by their very nature, ABV simulations obviate the existence of self-interstitial atoms (SIA), which are companion to vacancies during defect production in the primary damage phase \cite{2012CNMWolfer}. Neglecting SIA (as well as mixed interstitial) involvement in solute transport can often be justified when interstitial diffusion is orders of magnitude faster than that of vacancies, and --as importantly-- occurs in a (quasi) one-dimensional manner. This results in a point defect imbalance when SIAs reach defect sinks on time scales that are much shorter than those associated with vacancy motion, leaving vacancies as the sole facilitators of atomic transport \cite{1995PMLDuparc, 1996AMSoisson}. 
However, in certain cases interstitials play an important role in mediating solute diffusion, and their effect can no longer be dismissed when formulating global energy models for solute transport. A case in point is the recent discovery of solute drag by so-called `bridge' interstitial configurations in W-Re/Os alloys \cite{2015JNMSuzudo}, although several other examples exist \cite{2005JNMWillaime, 2015JNMGharaee, 2015SPRSetyawan}. In such cases, the ABV Hamiltonian is insufficient to capture the contribution of SIAs to microstructural evolution. This has prompted the development of cluster expansion Hamiltonians that include interstitials as well as vacancies as defect species \cite{2005PMSoisson, 2006JoNMSoisson, 2007PRBKrasnochtchekov, 2008JoNMVincent, 2010JoNMSoisson, 2012JoNMNgayam-Happy}. To date, however, an extension of such Hamiltonians to the Ising framework has not been attempted. That is the central objective of this paper.

Here, we propose a generalization of the ABV Ising model to ABVI systems of binary alloys subjected to irradiation. The paper is organized as follows: after this introduction, we describe our methodology in detail in Section \ref{theory}, providing a recipe to perform the ABV$\rightarrow$ABVI extension. Subsequently, in Section \ref{sec_results} we provide three different verification exercises in increasing order of complexity using published works. We finalize with a brief discussion and the conclusions in Section \ref{concl}.

\section{Theory and Numerical Methods}\label{theory}

\subsection{Cluster expansion Hamiltonians for binary alloys} \label{secH}

The most common approach to study the energetics of substitutional alloy systems is the cluster expansion method, in which the energies of the different crystal configurations are defined by specifying the occupation of each of the $N$ sites of a fixed crystal lattice by a number of distinct chemical species (which may include solvent and solute atoms, defects, etc.). This problem can quickly become intractable, given the combinatorial nature of arranging $N$ distinguishable atomic sites, and a number of approaches have been proposed to reduce the dimensional complexity of the problem \cite{1992PRBLaks, 1993PRBSanchez, 2004PRBBlum}. A common simplification is to assume that the Hamiltonian ${\cal H}$ of the system can be calculated as the sum of all possible pair interactions, defined by their \emph{bond} energies:
\begin{equation} 
{\cal H}= \sum_{\alpha,\beta} n_{\alpha\text{-}\beta} \epsilon_{\alpha\text{-}\beta}
\label{bonde}
\end{equation}
Where $\alpha$ and $\beta$ refer to a pair of lattice sites, $n$ is the total number of different bond types, and $\epsilon$ is the energy coefficients.

Further, a binary system containing two types of atoms (matrix) A and (solute) B, as well as vacancy defects is termed the `ABV' system, for which the pairwise cluster expansion Hamiltonian \eref{bonde} can be expressed as an Ising Hamiltonian of the following form \cite{1993PRBFrontera, 2000PASMaiAFratzl, 2003EELWeinkamer, 2004PTWeinkamer}:
\begin{equation} 
\label{habv}
{\cal H}= {\cal H}_0 + K\sum^{nn}_{\left<i,j\right>}\sigma^2_i\sigma^2_j + U\sum^{nn}_{\left<i,j\right>}\left(\sigma^2_i\sigma_j + \sigma^2_j\sigma_i\right) + J\sum^{nn}_{\left<i,j\right>}\sigma_i\sigma_j
\end{equation}
where $\langle i,j\rangle$ refers to a pair of lattice sites $i$ and $j$, and $\sigma$ are the occupancy variables:
\begin{equation}
\sigma=
\left\{\begin{array}{ll}
~~1 & \text{A (matrix atom)} \\ 
~~0 & \text{V (vacancy)} \\
-1 & \text{B (solute atom)}
\end{array}
\right.
\end{equation}
${\cal H}_0$ in eq.\ \eref{habv} is a constant independent of the configuration of lattice sites. The three coefficients $K$, $U$, and $J$  are:
\begin{eqnarray*}
K=& \nicefrac{1}{4}\left(\epsilon_{\text{A-A}} + \epsilon_{\text{B-B}} + 2\epsilon_{\text{A-B}}\right) + \left(\epsilon_{\text{V-V}} - \epsilon_{\text{A-V}} - \epsilon_{\text{B-V}}\right) \\
U=& \nicefrac{1}{4}\left(\epsilon_{\text{A-A}} - \epsilon_{\text{B-B}}\right) - \nicefrac{1}{2}\left(\epsilon_{\text{A-V}} - \epsilon_{\text{\text{B-V}}}\right) \\
J=& \nicefrac{1}{4}\left(\epsilon_{\text{A-A}} + \epsilon_{\text{B-B}} - 2\epsilon_{\text{A-B}}\right)
\end{eqnarray*}
These constants govern the kinetic behavior of the ABV system. The second term in the r.h.s.~of eq.\ \eref{habv} gives the relative importance of vacancies in the system. A large value of $K$ implies low vacancy concentrations, which in the limit of one single vacancy in the crystal converges to a constant value of $K'z\left(\nicefrac{N}{2} -1\right)$, where $K'=\nicefrac{1}{4}\left(\epsilon_{\text{A-A}}+\epsilon_{\text{B-B}}+2\epsilon_{\text{A-B}}\right)-
\left(\epsilon_{\text{A-V}}+\epsilon_{\text{B-V}}\right)$, and $z$ is the coordination number \cite{2004PTWeinkamer}. The asymmetry factor $U$ determines whether there is more affinity between A atoms and vacancies or B atoms and vacancies. $U>0$ indicates a preference of A-V pairs. $J$ determines the thermodynamics of the system, with $J>0$ leading to an ordered solid solution, $J<0$ to a phase-separated system, and $J=0$ resulting in an ideal solid solution. This Hamiltonian can be trivially extended from 1$^{\rm st}$ nearest neighbors ($nn$) to higher $nn$ by summing over all contributions:
\begin{equation}
{\cal H}= {\cal H}_{{\rm 1}^{\rm st}\text{-}nn} + {\cal H}_{{\rm 2}^{\rm nd}\text{-}nn} + \cdots
\end{equation}

\subsection{Generalization of the ABV Ising Hamiltonian to systems with interstitial atoms}

Next, we expand eq.\ \eref{habv} to a system containing A and B atoms, vacancies, and interstitial atoms, which we term `ABVI'. Interstitial atoms can be one of three distinct types, but in all cases two (otherwise substitutional) atoms share a single lattice position: AA denotes a self-interstitial atom (SIA), AB represents a mixed interstitial, and BB is a pure solute interstitial. Adding these extra species to the cluster expansion Hamiltonian brings the total number of species to six, which results in the following expression: 
\begin{equation}
\label{hinl}
{\cal H} = \sum_{\left< i,j \right>}^{nn} \sum_{\alpha, \beta} \epsilon_{\alpha\text{-}\beta}\lambda_i^{\alpha}\lambda_j^{\beta}\\
\end{equation}
where  $\alpha, \beta={\rm A, B, V, AA, AB, BB}$ and the occupancy variable $\lambda_i^{\alpha}=1$ if lattice site $i$ is occupied by type $\alpha$ and zero otherwise. The total number of independent terms in eq.\ \eref{hinl} is 36. However, assuming that a pair vacancy-interstitial is unstable up to several nearest neighbor distances, we can eliminate all the $\epsilon_{\text{V-I}}\lambda^{\text{V}}\lambda^{\text{I}}$ (where I= AA, AB, BB) terms in the equation, thus reducing the total number of terms to 30.

In the spirit of the ABV Ising model, we assign spin variables of different types to each of the species of the Hamiltonian:
\begin{equation}
\label{sabvi}
\sigma=
\left\{\begin{array}{ll}
~~2 & \text{AA (self-interstitial atom)} \\ 
~~1 & \text{A (matrix atom)} \\ 
~~0 & \text{V (vacancy) and AB (mixed interstitial)} \\
-1 & \text{B (solute atom)}\\
-2 & \text{BB (solute-solute interstitial)}
\end{array}
\right.
\end{equation}
Although the set of spin variables for the ABVI model is not unique, the one chosen above uses the lowest-order integer possible and preserves the \emph{magnetization} of the Ising model, {\it i.e.}~the excess amount of solvent after the solute has been subtracted out.
The convenience of choosing a zero spin variable for both the V and AB species brings about some complications in the Hamiltonian, however, which will be dealt with in Section \ref{correc}.

From their definition in eq.\ \eref{hinl}, the six independent $\lambda^{\alpha}$  variables can be written in terms of the spin variables furnished in eq.\ \eref{sabvi}:
\begin{eqnarray}
\label{lins}
\lambda^{\text{AA}} &= \frac{1}{24} \left( \sigma^{4} +2\sigma^{3} -\sigma^{2} -2\sigma \right)\nonumber \\
\lambda^{\text{A}} &= \frac{1}{6} \left( -\sigma^{4} -\sigma^{3} +4\sigma^{2} +4\sigma \right)\nonumber \\
\lambda^{\text{V}}&= \lambda^{AB} =\frac{1}{4} \left( \sigma^{4} -5\sigma^{2} +4 \right) \\
\lambda^{\text{B}} &= \frac{1}{6} \left( -\sigma^{4} +\sigma^{3} +4\sigma^{2} -4\sigma \right)\nonumber \\
\lambda^{\text{BB}} &= \frac{1}{24} \left( \sigma^{4} -2\sigma^{3} -\sigma^{2} +2\sigma \right)\nonumber
\end{eqnarray}
Inserting the above expressions into eq.\ \eref{hinl} and operating, the cluster expansion Hamiltonian is transformed into a generalized Ising system with integer spins:
\begin{eqnarray}
\label{hins}
{\cal H}&=& \sum_{\left< i,j \right>} \Big[C_{44}\sigma_i^4\sigma_j^4
+C_{43}\left(\sigma_i^4\sigma_j^3+\sigma_i^3\sigma_j^4\right)
+C_{42}\left(\sigma_i^4\sigma_j^2+\sigma_i^2\sigma_j^4\right)+\nonumber\\
&&+C_{41}\left(\sigma_i^4\sigma_j+\sigma_i\sigma_j^4\right) 
+C_{33}\sigma_i^3\sigma_j^3+C_{32}\left(\sigma_i^3\sigma_j^2+\sigma_i^2\sigma_j^3\right)+\nonumber\\
&&+C_{31}\left(\sigma_i^3\sigma_j+\sigma_i\sigma_j^3\right) 
+C_{22}\sigma_i^2\sigma_j^2
+C_{21}\left(\sigma_i^2\sigma_j+\sigma_i\sigma_j^2\right)+\nonumber\\
&&+C_{11}\sigma_i\sigma_j 
+C_{40}\left(\sigma_i^4+\sigma_j^4\right)
+C_{30}\left(\sigma_i^3+\sigma_j^3\right)+\nonumber\\
&&+C_{20}\left(\sigma_i^2+\sigma_j^2\right)
+C_{10}\left(\sigma_i+\sigma_j\right)+C_{00} \Big]
\end{eqnarray}
where $C_{mn}$ are the coefficients of the cluster expansion.

\subsubsection{Corrections to the Hamiltonian to separate V and AB contributions.}
\label{correc}

By construction, both vacancies and AB interstitials share $\sigma=0$ in eq.\ \eref{hins}, which in turn makes $\lambda_{\text{V}}=\lambda_{\text{AB}}=1$ leading to miscounting of both contributions. Corrections must therefore be adopted to recover the correct energy from the Hamiltonian. These corrections can simply be subtracted from the uncorrected Hamiltonian in eq.\ \eref{hins} as:
\begin{equation}
\label{hcorr}
{\cal H}_{\text{corrected}}= {\cal H}_{\text{uncorrected}} - [\text{correction terms}]
\end{equation}
The correction terms can be readily identified on inspection of eq.\ \eref{bonde}:
\begin{eqnarray}
\label{corr}
[\text{correction terms}]&=\epsilon_{\text{V-V}}n_{\text{AB-AB}} +\epsilon_{\text{AB-AB}}n_{\text{V-V}} +\epsilon_{\text{A-V}}n_{\text{A-AB}}+\nonumber\\
&+\epsilon_{\text{V-B}}n_{\text{AB-B}} +\epsilon_{\text{A-AB}}n_{\text{A-V}} +\epsilon_{\text{AB-B}}n_{\text{V-B}}
\end{eqnarray}
where $n_{\alpha\text{-}\beta}$ is numbers of bonds. Tracking the number of bonds in simulations takes extra computational effort, and also implies deviating from a purely Ising treatment. It is thus desirable to express $n_{\text{AB-AB}}$, $n_{\text{V-V}}$, $n_{\text{A-AB}}$, $n_{\text{AB-B}}$, $n_{\text{A-V}}$, and $n_{\text{V-B}}$ as summations of powers of the spin variables, as in eq.\ \eref{hins}. In this fashion, the correction terms do not add any additional cost to the evaluation of the Hamiltonian but, instead, only alter the value of the coefficients in eq.\ \eref{hins}. First, however, we must obtain expressions for all $n_{\alpha\text{-}\beta}$ in terms of the spin variable $\sigma$.

After discounting the $n_{\text{V-I}}$ terms (with I=AA, AB, BB), there are 18 $n_{\alpha\text{-}\beta}$ and therefore 18 independent equations are needed. 10 of them can be obtained from the summations of $\sigma$-polynomials:
\begin{eqnarray}
\label{nbonds_ssums1}
\sum\sigma_i\sigma_j&= 4n_{\text{AA-AA}}+2n_{\text{AA-A}}-2n_{\text{AA-B}}-4n_{\text{AA-BB}}+n_{\text{A-A}}+\nonumber\\
&-n_{\text{A-B}}-2n_{\text{A-BB}}+n_{\text{B-B}}+2n_{\text{B-BB}}+4n_{\text{BB-BB}}
\end{eqnarray}
\begin{eqnarray}
\sum \sigma_i^2\sigma_j+\sigma _i\sigma_j^2 &= 16n_{\text{AA-AA}}+6n_{\text{AA-A}}-2n_{\text{AA-B}}+2n_{\text{A-A}}+2n_{\text{A-BB}}+\nonumber\\
&-2n_{\text{B-B}}-6n_{\text{B-BB}}-16n_{\text{BB-BB}}
\end{eqnarray}
\begin{eqnarray}
\sum \sigma_i^2\sigma _j^2 &= 16n_{\text{AA-AA}}+4n_{\text{AA-A}}+4n_{\text{AA-B}}+16n_{\text{AA-BB}}+ n_{\text{A-A}}+\nonumber\\
&+n_{\text{A-B}}+4n_{\text{A-BB}}+n_{\text{B-B}}+4n_{\text{B-BB}}+16n_{\text{BB-BB}}
\end{eqnarray}
\begin{eqnarray}
\sum \sigma_i^3\sigma_j+\sigma _i\sigma_j^3& = 32n_{\text{AA-AA}}+10n_{\text{AA-A}}-10n_{\text{AA-B}}-32n_{\text{AA-BB}}+\nonumber\\
&+2n_{\text{A-A}}-2n_{\text{A-B}}-10n_{\text{A-BB}}+2n_{\text{B-B}}+10n_{\text{B-BB}}+\nonumber\\
&+32n_{\text{BB-BB}}
\end{eqnarray}
\begin{eqnarray}
\sum \sigma_i^3\sigma_j^2+\sigma _i^2\sigma_j^3 &= 64n_{\text{AA-AA}}+12n_{\text{AA-A}}+4n_{\text{AA-B}}+ 2n_{\text{A-A}}-4n_{\text{A-BB}}+\nonumber\\
&-2n_{\text{B-B}}-12n_{\text{B-BB}}-64n_{\text{BB-BB}}
\end{eqnarray}
\begin{eqnarray}
\sum \sigma_i^3\sigma_j^3 &= 64n_{\text{AA-AA}}+8n_{\text{AA-A}}-8n_{\text{AA-B}}-64n_{\text{AA-BB}}+ n_{\text{A-A}}-n_{\text{A-B}}+\nonumber\\
&-8n_{\text{A-BB}}+n_{\text{B-B}}-8n_{\text{B-BB}}+64n_{\text{BB-BB}}
\end{eqnarray}
\begin{eqnarray}
\sum \sigma_i^4\sigma_j+\sigma _i\sigma_j^4 &= 64n_{\text{AA-AA}}+18n_{\text{AA-A}}-14n_{\text{AA-B}}+2n_{\text{A-A}}+\nonumber\\
&+14n_{\text{A-BB}}-2n_{\text{B-B}}-18n_{\text{B-BB}}-64n_{\text{BB-BB}}
\end{eqnarray}
\begin{eqnarray}
\sum \sigma_i^4\sigma_j^2+\sigma_i^2\sigma_j^4&= 128n_{\text{AA-AA}}+20n_{\text{AA-A}}+20n_{\text{AA-B}}+128n_{\text{AA-BB}}+\nonumber\\
&+2n_{\text{A-A}}+2n_{\text{A-B}}+20n_{\text{A-BB}}+2n_{\text{B-B}}+20n_{\text{B-BB}}+\nonumber\\
&+128n_{\text{BB-BB}}
\end{eqnarray}
\begin{eqnarray}
\sum \sigma_i^4\sigma_j^3+\sigma_i^3\sigma_j^4 &= 256n_{\text{AA-AA}}+24n_{\text{AA-A}}-8n_{\text{AA-B}}+2n_{\text{A-A}}+\nonumber\\
&+8n_{\text{A-BB}}-2n_{\text{B-B}}-24n_{\text{B-BB}}-256n_{\text{BB-BB}}
\end{eqnarray}
\begin{eqnarray}
\sum \sigma_i^4\sigma_j^4 &= 256n_{\text{AA-AA}}+16n_{\text{AA-A}}+16n_{\text{AA-B}}+256n_{\text{AA-BB}}+ n_{\text{A-A}}+\nonumber\\
&+n_{\text{A-B}}+16n_{\text{A-BB}}+n_{\text{B-B}}+16n_{\text{B-BB}}+256n_{\text{BB-BB}}
\label{nbonds_ssums2}
\end{eqnarray}
However, the above equations do not contain any $n_{\alpha\text{-}\beta}$ with $\alpha$ or $\beta$= V, AB. Six more equations that do contain these terms can be obtained by counting the numbers of six species $N_{\alpha}$:
\begin{eqnarray} 
\label{nbonds_natoms1}
zN_{\text{AA}} &= 2n_{\text{AA-AA}}+n_{\text{AA-A}}+n_{\text{AA-AB}}+n_{\text{AA-B}}+n_{\text{AA-BB}} \\
zN_{\text{A}} &= n_{\text{AA-A}}+2n_{\text{A-A}}+n_{\text{A-V}}+n_{\text{A-AB}}+n_{\text{A-B}}+n_{\text{A-BB}} \\
zN_{\text{V}} &= n_{\text{A-V}}+2n_{\text{V-V}}+n_{\text{V-B}} \\
zN_{\text{AB}} &= n_{\text{AA-AB}}+n_{\text{A-AB}}+2n_{\text{AB-AB}}+n_{\text{AB-B}}+n_{\text{AB-BB}} \\
zN_{\text{B}} &= n_{\text{AA-B}}+n_{\text{A-B}}+n_{\text{V-B}}+n_{\text{AB-B}}+2n_{\text{B-B}}+n_{\text{B-BB}} \\
zN_{\text{BB}} &= n_{\text{AA-BB}}+n_{\text{A-BB}}+n_{\text{AB-BB}}+n_{\text{B-BB}}+2n_{\text{BB-BB}}
\label{nbonds_natoms2}
\end{eqnarray}
where $z$ is the coordination number. Combining eqs.\ \eref{nbonds_ssums1} through \eref{nbonds_natoms2}, we have 
16 equations with 18 unknowns. In order to solve the system, we express everything parametrically in terms of two bond numbers, $n_{\text{AB-A}}$ and $n_{\text{AB-B}}$\footnote{This choice is justified both by the fact that neither A-AB nor AB-B bonds are very likely to appear in the simulations, and because --as will pointed out below-- AB interstitialcy jumps are the likeliest to change the global concentration of species, which results in the need to update the non-configurational constants in the ABVI Hamiltonian (cf.\ eq.\ \eref{nonconf}).}, and solve for the rest of the $n_{\alpha\text{-}\beta}$. $n_{\text{AB-A}}$ and $n_{\text{AB-B}}$ are then the only bond numbers that must be calculated on the fly in  the kMC simulations.

\subsubsection{The corrected Ising Hamiltonian.}
\label{corrected}

After solving for all $n_{\alpha\text{-}\beta}$, the corrected Hamiltonian can be obtained by substituting eq.\ \eref{corr} into eq.\ \eref{hcorr}. Except for an additional term $C_0$, the final expression of the corrected Hamiltonian is the same as the uncorrected one in eq.\ \eref{hins}. However, the coefficients $C_{mn}$ are now `corrected' to account for the AB/V conflict. Based on the physical characteristics of each coefficient, each term in the Hamiltonian of the ABVI system can be grouped into three different configurational classes and one non-configurational group:
\begin{eqnarray}
\label{hclass}
{\cal H}_{\text{corrected}}&= \sum_{\left< i,j \right>}^{nn} \left[C_{44}\sigma_i^4\sigma_j^4+C_{42}\left(\sigma_i^4\sigma_j^2+\sigma_i^2\sigma_j^4\right)
+C_{22}      \sigma_i^2\sigma_j^2\right]+~\textcolor{blue}{\left(\text{Class 1}\right)}
\nonumber\\
&+\sum_{\left< i,j \right>}^{nn}\Big[C_{43}\left(\sigma_i^4\sigma_j^3+\sigma_i^3\sigma_j^4\right)+C_{41}\left(\sigma_i^4\sigma_j+\sigma_i\sigma_j^4\right)+C_{32}\left(\sigma_i^3\sigma_j^2+\sigma_i^2\sigma_j^3\right)+\nonumber\\
&\quad\quad\quad\quad+C_{21}\left(\sigma_i^2\sigma_j+\sigma_i\sigma_j^2\right)\Big]+~\textcolor{blue}{\left(\text{Class 2}\right)}
\nonumber\\
&+\sum_{\left< i,j \right>}^{nn} \Big[C_{33}      \sigma_i^3\sigma_j^3
+C_{31}\left(\sigma_i^3\sigma_j+\sigma_i\sigma_j^3\right)
+C_{11}      \sigma_i\sigma_j
\Big]+ \
\textcolor{blue}{\left(\text{Class 3}\right)}\nonumber\\
&+\sum_{\left< i,j \right>}^{nn} \left[\right.
C_{40}\left(\sigma_i^4+\sigma_j^4\right)
+C_{30}\left(\sigma_i^3+\sigma_j^3\right)
+C_{20}\left(\sigma_i^2+\sigma_j^2\right)+\nonumber\\
&\quad\quad\quad\quad+C_{10}\left(\sigma_i+\sigma_j\right)
+C_{00}  
\left.\right]+ C_0 \label{nonconf}
\textcolor{blue}~{\left(\textcolor{blue}{\text{Non-configurational}}\right)}
\end{eqnarray}
where the coefficients $C_{mn}$ are:

\textcolor{blue}{Class 1}
\begin{eqnarray*} 
C_{44} &=  \frac{1}{576}\Big\{\left(\epsilon_{\text{AA-AA}}-8\epsilon_{\text{AA-A}}-8\epsilon_{\text{AA-B}}+2\epsilon_{\text{AA-BB}}-8\epsilon_{\text{A-BB}}-8\epsilon_{\text{B-BB}}+\epsilon_{\text{BB-BB}}\right)+\\  
&+\left(12\epsilon_{\text{AA-AB}}-12\epsilon_{\text{AB-AB}}+12\epsilon_{\text{AB-BB}}\right)+\left(-48\epsilon_{\text{A-V}}+48\epsilon_{\text{V-V}}-48\epsilon_{\text{V-B}}\right)+\\
&+\left(16\epsilon_{\text{A-A}}+32\epsilon_{\text{A-B}}+16\epsilon_{\text{B-B}}\right)\Big\}\\
C_{42} &= \frac{1}{576}\Big\{\left(-\epsilon_{\text{AA-AA}}+20\epsilon_{\text{AA-A}}+20\epsilon_{\text{AA-B}}-2\epsilon_{\text{AA-BB}}+20\epsilon_{\text{A-BB}}+20\epsilon_{\text{B-BB}}-\epsilon_{\text{BB-BB}}\right)+\\ 
&+\left(-36\epsilon_{\text{AA-AB}}+36\epsilon_{\text{AB-AB}}-36\epsilon_{\text{AB-BB}}\right)+\left(216\epsilon_{\text{A-V}}-216\epsilon_{\text{V-V}}+216\epsilon_{\text{V-B}}\right)+\\
& +\left(-64\epsilon_{\text{A-A}}-128\epsilon_{\text{A-B}}-64\epsilon_{\text{B-B}}\right)\Big\}\\
C_{22} &= \frac{1}{576}\Big\{\left(\epsilon_{\text{AA-AA}}-32\epsilon_{\text{AA-A}}-32\epsilon_{\text{AA-B}}+2\epsilon_{\text{AA-BB}}-32\epsilon_{\text{A-BB}}-32\epsilon_{\text{B-BB}}+\epsilon_{\text{BB-BB}}\right)+\\
& +\left(60\epsilon_{\text{AA-AB}}-60\epsilon_{\text{AB-AB}}+60\epsilon_{\text{AB-BB}}\right)+\left(-960\epsilon_{\text{A-V}}+960\epsilon_{\text{V-V}}-960\epsilon_{\text{V-B}}\right)+\\
& +\left(256\epsilon_{\text{A-A}}+512\epsilon_{\text{A-B}}+256\epsilon_{\text{B-B}}\right)\Big\}
\end{eqnarray*}

\textcolor{blue}{Class 2}
\begin{eqnarray*}
C_{43} &= \frac{1}{288}\Big\{\left(\epsilon_{\text{AA-AA}}-6\epsilon_{\text{AA-A}}-2\epsilon_{\text{AA-B}}+2\epsilon_{\text{A-BB}}+6\epsilon_{\text{B-BB}}-\epsilon_{\text{BB-BB}}\right)+\\ 
& +\left(6\epsilon_{\text{AA-AB}}-6\epsilon_{\text{AB-BB}}\right)+\left(-12\epsilon_{\text{A-V}}+12\epsilon_{\text{V-B}}\right)+\left(8\epsilon_{\text{A-A}}-8\epsilon_{\text{B-B}}\right)\Big\}\\
C_{41} &=
\frac{1}{288}\Big\{\left(-\epsilon_{\text{AA-AA}}+12\epsilon_{\text{AA-A}}-4\epsilon_{\text{AA-B}}+4\epsilon_{\text{A-BB}}-12\epsilon_{\text{B-BB}}+\epsilon_{\text{BB-BB}}\right)+\\
& +\left(-6\epsilon_{\text{AA-AB}}+6\epsilon_{\text{AB-BB}}\right)+\left(48\epsilon_{\text{A-V}}-48\epsilon_{\text{V-B}}\right)+\left(-32\epsilon_{\text{A-A}}+32\epsilon_{\text{B-B}}\right)\Big\}\\
C_{32} &=
\frac{1}{288}\Big\{\left(-\epsilon_{\text{AA-AA}}+18\epsilon_{\text{AA-A}}+14\epsilon_{\text{AA-B}}-14\epsilon_{\text{A-BB}}-18\epsilon_{\text{B-BB}}+\epsilon_{\text{BB-BB}}\right)+\\ 
& +\left(-30\epsilon_{\text{AA-AB}}+30\epsilon_{\text{AB-BB}}\right)+\left(60\epsilon_{\text{A-V}}-60\epsilon_{\text{V-B}}\right)+\left(-32\epsilon_{\text{A-A}}+32\epsilon_{\text{B-B}}\right)\Big\}\\
C_{21} &=
\frac{1}{288}\Big\{\left(\epsilon_{\text{AA-AA}}-24\epsilon_{\text{AA-A}}-8\epsilon_{\text{AA-B}}+8\epsilon_{\text{A-BB}}+24\epsilon_{\text{B-BB}}-\epsilon_{\text{BB-BB}}\right)+\\ 
& +\left(30\epsilon_{\text{AA-AB}}-30\epsilon_{\text{AB-BB}}\right)+\left(-240\epsilon_{\text{A-V}}+240\epsilon_{\text{V-B}}\right)+\left(128\epsilon_{\text{A-A}}-128\epsilon_{\text{B-B}}\right)\Big\}
\end{eqnarray*}

\textcolor{blue}{Class 3}
\begin{eqnarray*}
C_{33} &=
\frac{1}{144}\Big\{\left(\epsilon_{\text{AA-AA}}-4\epsilon_{\text{AA-A}}+4\epsilon_{\text{AA-B}}-2\epsilon_{\text{AA-BB}}+4\epsilon_{\text{A-BB}}-4\epsilon_{\text{B-BB}}+\epsilon_{\text{BB-BB}}\right)+
\\ & +
\left(4\epsilon_{\text{A-A}}-8\epsilon_{\text{A-B}}+4\epsilon_{\text{B-B}}\right)\Big\}
\\
C_{31} &=
\frac{1}{144}\Big\{\left(-\epsilon_{\text{AA-AA}}+10\epsilon_{\text{AA-A}}-10\epsilon_{\text{AA-B}}+2\epsilon_{\text{AA-BB}}-10\epsilon_{\text{A-BB}}+10\epsilon_{\text{B-BB}}-\epsilon_{\text{BB-BB}}\right)+
\\ & +
\left(-16\epsilon_{\text{A-A}}+32\epsilon_{\text{A-B}}-16\epsilon_{\text{B-B}}\right)\Big\}
\\
C_{11} &=
\frac{1}{144}\Big\{\left(\epsilon_{\text{AA-AA}}-16\epsilon_{\text{AA-A}}+16\epsilon_{\text{AA-B}}-2\epsilon_{\text{AA-BB}}+16\epsilon_{\text{A-BB}}-16\epsilon_{\text{B-BB}}+\epsilon_{\text{BB-BB}}\right)+
\\ & +
\left(64\epsilon_{\text{A-A}}-128\epsilon_{\text{A-B}}+64\epsilon_{\text{B-B}}\right)\Big\}
\end{eqnarray*}

\textcolor{blue}{Non-Configurational}
\begin{eqnarray*}
C_{40} &=
\frac{1}{24}\Big\{\left(\epsilon_{\text{AA-AB}}+\epsilon_{\text{AB-BB}}\right)+\left(-4\epsilon_{\text{A-AB}}+6\epsilon_{\text{AB-AB}}-4\epsilon_{\text{AB-B}}\right)+
\\ & +
\left(-4\epsilon_{\text{A-V}}+6\epsilon_{\text{V-V}}-4\epsilon_{\text{V-B}}\right)\Big\}
\\
C_{30} &=
\frac{1}{12}\Big\{\left(\epsilon_{\text{AA-AB}}-\epsilon_{\text{AB-BB}}\right)+\left(-2\epsilon_{\text{A-AB}}+2\epsilon_{\text{AB-B}}\right)+\left(-2\epsilon_{\text{A-V}}+2\epsilon_{\text{V-B}}\right)\Big\}
\\
C_{20} &=
\frac{1}{24}\Big\{\left(-\epsilon_{\text{AA-AB}}-\epsilon_{\text{AB-BB}}\right)+\left(16\epsilon_{\text{A-AB}}-30\epsilon_{\text{AB-AB}}+16\epsilon_{\text{AB-B}}\right)+
\\ & +
\left(16\epsilon_{\text{A-V}}-30\epsilon_{\text{V-V}}+16\epsilon_{\text{V-B}}\right)\Big\}
\\
C_{10} &=
\frac{1}{12}\Big\{\left(-\epsilon_{\text{AA-AB}}+\epsilon_{\text{AB-BB}}\right)+\left(8\epsilon_{\text{A-AB}}-8\epsilon_{\text{AB-B}}\right)+\left(8\epsilon_{\text{A-V}}-8\epsilon_{\text{V-B}}\right)\Big\}
\\
C_{00} &=
\left(\epsilon_{\text{AB-AB}}+\epsilon_{\text{V-V}}\right)
\\
C_{0} &=
\frac{n_{\text{A-AB}}}{2}\left(-\epsilon_{\text{AB-AB}}+2\epsilon_{\text{A-AB}}-2\epsilon_{\text{A-V}}+\epsilon_{\text{V-V}}\right)+
\\ & +
\frac{n_{\text{AB-B}}}{2}\left(-\epsilon_{\text{AB-AB}}+2\epsilon_{\text{AB-B}}-2\epsilon_{\text{V-B}}+\epsilon_{\text{V-V}}\right)+
\\ & +
\frac{Z}{2}\left[N_{A}\left(\epsilon_{\text{AB-AB}}-2\epsilon_{\text{A-AB}}\right)+N_{B}\left(\epsilon_{\text{AB-AB}}-2\epsilon_{\text{AB-B}}\right)\right]+
\\ & -
\frac{Z}{2}\left[N_{V}\epsilon_{\text{AB-AB}}+N_{AA}\epsilon_{\text{V-V}}-N_{AB}\epsilon_{\text{V-V}}+N_{BB}\epsilon_{\text{V-V}}\right]
\end{eqnarray*}
This way of grouping the $C_{mn}$ is not unique. We have chosen the three classes above to represent a given physical behavior along the lines of the coefficients $K$, $U$, $J$ of the ABV Ising model. Loosely speaking, the physical meanings of each of the three classes is as follows:
\begin{itemize}
\item Class 1 (even-even power terms) gives the relative importance of interactions between point defects (vacancies and interstitials).
\item Class 2 (even-odd power terms) gives the affinity between atoms and point defects.
\item Class 3 (odd-odd power terms) determines the equilibrium phase diagram.
\end{itemize}
In the standard ABV model, defect (vacancy) hops do not change the global species concentrations. That means that the \emph{non-configurational} class of terms in the Hamiltonian \eref{hclass} does not change merely by vacancy jumps. However, the ABVI model now allows for defect transitions that change the global balance of species\footnote{The most obvious one being a vacancy-interstitial recombination.}. Specifically, there are two types of transitions that affect the species concentrations when they occur. The first one involves vacancy-interstitial recombinations:
\begin{eqnarray*}
\text{AA+V $\rightarrow$ A + A} \\
\text{AB+V $\rightarrow$ A + B} \\
\text{BB+V $\rightarrow$ B + B}
\end{eqnarray*}
The second type is related to the \emph{interstitialcy} mechanism, by which an interstitial atom displaces an atom from an adjacent lattice position so that it becomes the interstitial in its turn, able to displace another atom. This mechanism includes four reactions:
\begin{eqnarray*}
\text{AA + B $\rightarrow$ A + AB} \\
\text{AB + A $\rightarrow$ B + AA} \\
\text{AB + B $\rightarrow$ A + BB} \\
\text{BB + A $\rightarrow$ B + AB}
\end{eqnarray*}
Except when one of the above reactions occurs, the incremental energy formulation used to compute energy differences between the initial and final states allows us to discard the non-configurational terms during calculations.

In order to truly represent a generalized Hamiltonian, the ABVI model Hamiltonian must reduce to the AV and ABV models in their respective limits (AV: no solute, vacancies; ABV: solute plus vacancies). Indeed, we have conducted verification tests of both particular cases and we have found matching results. This is the subject of Sec.\ \ref{sec_results}, where we have simulated the time evolution of ABV and ABVI systems using the generalized Hamiltonian presented above. Our method of choice is kinetic Monte Carlo (kMC), which we describe in detail in the following section.

\subsection{Kinetic Monte Carlo Simulation}

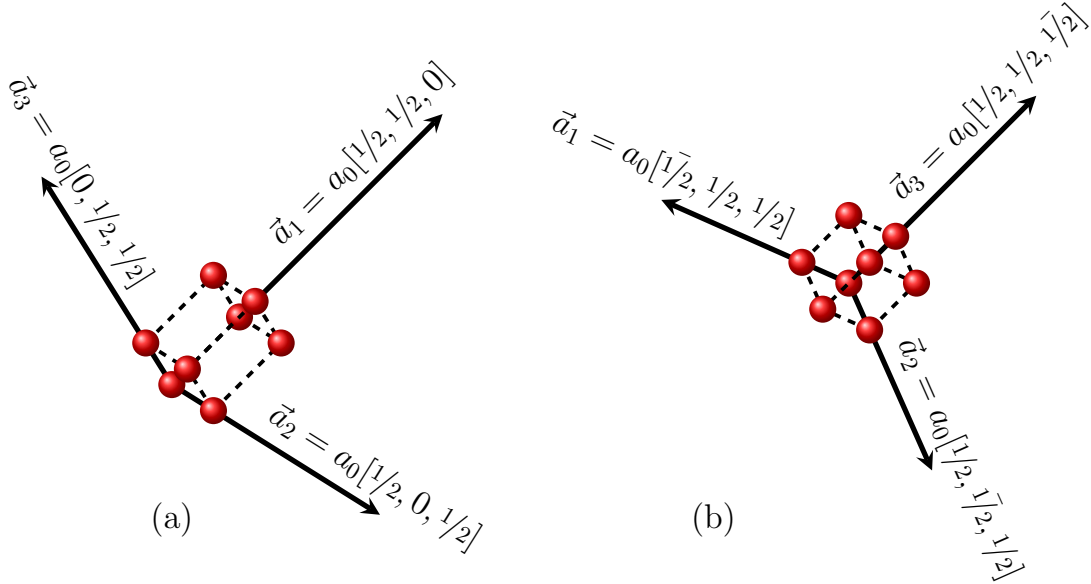
\begin{figure}[h]
	\centering
		\begin{tikzpicture}[scale=0.9, every node/.style={scale=0.9}]
	\draw[dashed, line width=0.5mm] (0, 0, 0) -- (1, 1, 0);
	\draw[dashed, line width=0.5mm] (0, 0, 0) -- (1, 0, 1);
	\draw[dashed, line width=0.5mm] (0, 0, 0) -- (0, 1, 1);
	
	\draw[dashed, line width=0.5mm] (1, 1, 0) -- (1, 2, 1);
	\draw[dashed, line width=0.5mm] (1, 1, 0) -- (2, 1, 1);
	
	\draw[dashed, line width=0.5mm] (1, 0, 1) -- (2, 1, 1);
	\draw[dashed, line width=0.5mm] (1, 0, 1) -- (1, 1, 2);

	\draw[dashed, line width=0.5mm] (0, 1, 1) -- (1, 2, 1);
	\draw[dashed, line width=0.5mm] (0, 1, 1) -- (1, 1, 2);
	
	\draw[->, >=stealth, line width=0.7mm] (1, 1, 0) -- (4, 4, 0) node [pos= 0.7, above, sloped, font=\large] {$\vec{a}_1=a_0 [\nicefrac{1}{2},\nicefrac{1}{2},0]$};
	\draw[->, >=stealth, line width=0.7mm] (0, 0, 0) -- (5, 0, 5) node [pos= 0.9, above, sloped, font=\large] {$\vec{a}_2= a_0[\nicefrac{1}{2},0,\nicefrac{1}{2}]$};
	\draw[->, >=stealth, line width=0.7mm] (0, 0, 0) -- (0, 5, 5) node [pos= 0.9, above, sloped, font=\large] {$\vec{a}_3= a_0[0,\nicefrac{1}{2},\nicefrac{1}{2}]$};
	
	\shade[ball color= red] (0,0,0) circle (0.2cm);
	\shade[ball color= red] (1, 1, 0) circle (0.2cm);
	\shade[ball color= red] (1, 0, 1) circle (0.2cm);
	\shade[ball color= red] (0, 1, 1) circle (0.2cm);
		\draw[dashed, line width=0.5mm] (2, 2, 2) -- (1, 1, 2);
		\draw[dashed, line width=0.5mm] (2, 2, 2) -- (1, 2, 1);
		\draw[dashed, line width=0.5mm] (2, 2, 2) -- (2, 1, 1);
	\shade[ball color= red] (1, 1, 2) circle (0.2cm);
	\shade[ball color= red] (1, 2, 1) circle (0.2cm);
	\shade[ball color= red] (2, 1, 1) circle (0.2cm);
	\shade[ball color= red] (2,2,2) circle (0.2cm);
	\node[font=\large] at (0, -2) {(a)};
	\node[font=\large] at (8, -2) {(b)};

\begin{scope}[shift={(10, 1.5)}]
	\draw[dashed, line width=0.5mm] (0, 0, 0) -- (-0.5,  0.5,  0.5);
	\draw[dashed, line width=0.5mm] (0, 0, 0) -- ( 0.5, -0.5,  0.5);
	\draw[dashed, line width=0.5mm] (0, 0, 0) -- ( 0.5,  0.5, -0.5);
	
	\draw[dashed, line width=0.5mm] (-0.5, 0.5, 0.5) -- (0, 0, 1);
	\draw[dashed, line width=0.5mm] (-0.5, 0.5, 0.5) -- (0, 1, 0);
	
	\draw[dashed, line width=0.5mm] ( 0.5,-0.5, 0.5) -- (1, 0, 0);
	\draw[dashed, line width=0.5mm] ( 0.5,-0.5, 0.5) -- (0, 0, 1);
	
	\draw[dashed, line width=0.5mm] ( 0.5, 0.5,-0.5) -- (1, 0, 0);
	\draw[dashed, line width=0.5mm] ( 0.5, 0.5,-0.5) -- (0, 1, 0);

	\draw[->, >=stealth, line width=0.7mm] (0, 0, 0) -- (-2,  2,  2) node [pos= 1, above, sloped, font=\large] {$\vec{a}_1= a_0[\bar{\nicefrac{1}{2}},\nicefrac{1}{2},\nicefrac{1}{2}]$};
	\draw[->, >=stealth, line width=0.7mm] (0, 0, 0) -- ( 2, -2,  2) node [pos= 1, above, sloped, font=\large] {$\vec{a}_2=a_0 [\nicefrac{1}{2},\bar{\nicefrac{1}{2}},\nicefrac{1}{2}]$};
	\draw[->, >=stealth, line width=0.7mm] (0, 0, -0) -- ( 2,  2, -2) node [pos= 0.85, above, sloped, font=\large] {$\vec{a}_3=a_0 [\nicefrac{1}{2},\nicefrac{1}{2},\bar{\nicefrac{1}{2}}]$};
	
	\shade[ball color= red] (0,0,0) circle (0.2cm);
		\draw[dashed, line width=0.5mm] ( 0.5, 0.5, 0.5) -- (1, 0, 0);
		\draw[dashed, line width=0.5mm] ( 0.5, 0.5, 0.5) -- (0, 1, 0);
		\draw[dashed, line width=0.5mm] ( 0.5, 0.5, 0.5) -- (0, 0, 1);
	\shade[ball color= red] (-0.5, 0.5, 0.5) circle (0.2cm);
	\shade[ball color= red] ( 0.5,-0.5, 0.5) circle (0.2cm);
	\shade[ball color= red] ( 0.5, 0.5,-0.5) circle (0.2cm);
	\shade[ball color= red] (0, 0, 1) circle (0.2cm);
	\shade[ball color= red] (0, 1, 0) circle (0.2cm);
	\shade[ball color= red] (1, 0, 0) circle (0.2cm);
	\shade[ball color= red] ( 0.5, 0.5, 0.5) circle (0.2cm);
\end{scope}

		\end{tikzpicture}
	\caption{Primitive cells for (a) FCC and (b) BCC lattices showing all eight vertices as red spheres. The vectors $\vec{a_1}$, $\vec{a_2}$, and $\vec{a_3}$ are primitive basis of crystal, with $a_0$ the lattice parameter.}
	\label{fig_rhombic}
\end{figure}

In this section we discuss relevant details of the kMC simulation method in relation to our extended ABVI model. All simulations are conducted on a rigid lattice generated from trigonal (primitive)
representations of face-centered cubic (FCC) and body-centered cubic (BCC) crystals. The primitive cells employed for each crystal structure are provided in Figure \ref{fig_rhombic}. The simulations are generally conducted in the grand canonical ensemble, to allow for irradiation damage simulations when required \cite{chestnut1963}. All kinetic transitions are assumed to be due to defect hops. In particular, we consider the vacancy and interstitialcy mechanisms to enable atomic transport. After every transition, the configuration of the system is updated and a new transition is considered.

\subsubsection{Residence-time algorithm.}
We use the residence-time algorithm (RTA) \cite{1966PPSYoung} to track the kinetic evolution of the system through a series of thermally activated transitions. The transition rates $R_{ij}$ connecting an initial state $i$ to a final state $j$ are calculated as:
\begin{equation} \label{rate}
	r_{ij}= \nu \exp \left(-\frac{\Delta E_{ij}}{k_BT} \right)
\end{equation}
where $\Delta E_{ij}>0$ is an activation energy that will be discussed below, $\nu$ is the attempt frequency, and $1/k_BT$ is the reciprocal temperature. With the system in configuration $i$, an event is randomly chosen with a probability proportional to its rate, and the time advanced per kMC step is on average $\delta t_i= \left(\sum_jr_{ij}\right)^{-1}$. In addition to thermally activated transitions such as those represented by eq.\ \eref{rate}, we consider spontaneous events --for which, strictly speaking, $\Delta E_{ij}$ may be negative-- such as recombination between vacancies and interstitials, absorption at sinks, etc. These events occur instantaneously with $\delta t=0$.

\subsubsection{Activation energy models.} \label{sec_actE}
There are several models proposed to describe the activation energy, which are based on different interpretations of the atomic migration process (see for e.g., \cite{2010JoNMSoisson} for a recent review). The first model is the so-called \emph{saddle-point energy} model (also known as `cut-bond' model in \cite{2008JoNMVincenta}) \cite{1996AMSoisson, 2002_PRBBouar, 2007PRBSoisson}. The activation energy is given by:
\begin{equation} \label{saddle}
	\Delta E_{ij}= E^{SP}_{XY} - \sum_n \epsilon_{X\text{-}n} - \sum_{p\neq X} \epsilon_{Y\text{-}p}
\end{equation}
where $Y$ refers to the defect (e.g.\ a vacancy), and $X$ to the atom exchanging positions with $Y$. The later two summations are the bonding energies between $X$, $Y$ and the adjacent neighbor sites $n$ and $p$. In this model, the energy barrier is calculated as the difference between the energy of the system at the saddle point and that of the initial state, symbolized by the two summations in the r.h.s.\ of eq.\ \eref{saddle}. These summations can be computed using the ABVI Ising Hamiltonian formulas described in Section \ref{secH}. The saddle-point energy $E^{SP}_{XY}$  is generally taken to be a constant \cite{1996AMSoisson}, or is computed as a especial sum of bond energies of the jumping atom at the saddle point: $E^{SP}_{XY}= \sum_q \epsilon^{SP, Y}_{Xq}$ \cite{2002_PRBBouar, 2007PRBSoisson}.

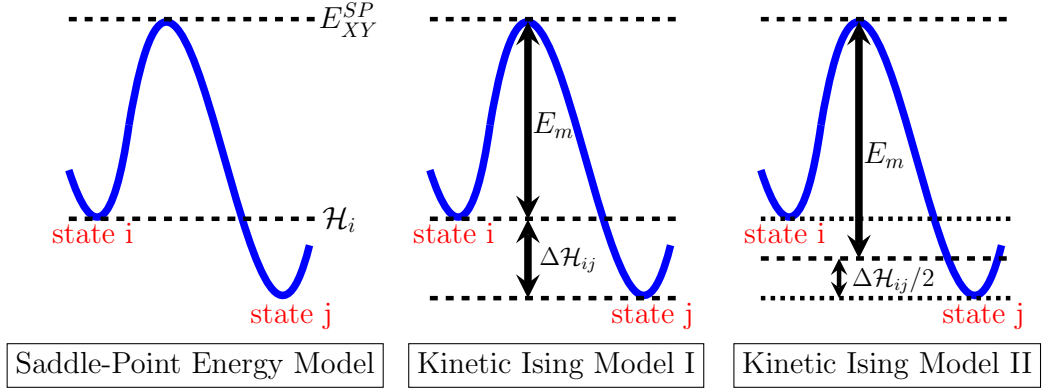
\begin{figure}[tbh]
	\centering
		\begin{tikzpicture}[scale=0.8, every node/.style={scale=0.8}]

\clip (-2, 2.5) rectangle (17, -5); 
		
	\draw[blue, line width=1mm] (0,-0.75) .. controls (0.4, -2) and (0.75,-1.75)  .. (1,0);
	\draw[blue, line width=1mm] (1,0) .. controls (2, 6) and (3, -6)  .. (4,-2);
	\draw[dashed, line width= 0.6mm] (0,  1.76) -- (4.2, 1.76);
	\draw[dashed, line width= 0.6mm] (0, -1.56) -- (4.2, -1.56);
	\node[red, font=\large] at (0.4, -1.8) {state i};
	\node[red, font=\large] at (3.7, -3.17) {state j};

	\node[draw, font=\large] at (2.1, -4) {Saddle-Point Energy Model};	
	\node[font=\large] at (4.65, 1.75) {$E^{SP}_{XY}$};
	\node[font=\large] at (4.5, -1.57) {${\cal H}_{i}$};
		
\begin{scope}[shift={(6, 0)}]
	\draw[blue, line width=1mm] (0,-0.75) .. controls (0.4, -2) and (0.75,-1.75)  .. (1,0);
	\draw[blue, line width=1mm] (1,0) .. controls (2, 6) and (3, -6)  .. (4,-2);
	\draw[dashed, line width= 0.6mm] (0,  1.76) -- (4.2, 1.76);
	\draw[dashed, line width= 0.6mm] (0, -1.56) -- (4.2, -1.56);
	\node[red, font=\large] at (0.4, -1.8) {state i};
	\node[red, font=\large] at (3.7, -3.17) {state j};

	\node[draw, font=\large] at (2.1, -4) {Kinetic Ising Model I};
	\draw[dashed, line width= 0.6mm] (0, -2.88) -- (4.2, -2.88);		
	\draw[<->, >=stealth, line width=1mm] (1.63,  1.72) -- (1.63, -1.57);
	\draw[<->, >=stealth, line width=1mm] (1.63, -1.57) -- (1.63, -2.88);
	\node[font=\large] at (2.05, 0) {$E_{m}$};
	\node at (2.3, -2.25) {$\Delta {\cal H}_{ij}$};
\end{scope}

\begin{scope}[shift={(11.5, 0)}]
	\draw[blue, line width=1mm] (0,-0.75) .. controls (0.4, -2) and (0.75,-1.75)  .. (1,0);
	\draw[blue, line width=1mm] (1,0) .. controls (2, 6) and (3, -6)  .. (4,-2);
	\draw[dashed, line width= 0.6mm] (0,  1.76) -- (4.2, 1.76);
	\draw[dotted, line width= 0.6mm] (0, -1.56) -- (4.2, -1.56); 
	\node[red, font=\large] at (0.4, -1.8) {state i};
	\node[red, font=\large] at (3.7, -3.17) {state j};

	\node[draw, font=\large] at (2.1, -4) {Kinetic Ising Model II};
	\draw[dashed, line width= 0.6mm] (0, -2.22) -- (4.2, -2.22);
	\draw[dotted, line width= 0.6mm] (0, -2.88) -- (4.2, -2.88);		
	\draw[<->, >=stealth, line width=1mm] (1.63,  1.72) -- (1.63, -2.22);
	\draw[<->, >=stealth, line width=0.6mm] (1.3, -2.22) -- (1.3, -2.88);
	\node[font=\large] at (2.05, -0.5) {$E_{m}$};
	\node at (2.2, -2.58) {$\Delta {\cal H}_{ij}/2$};
\end{scope}

		\end{tikzpicture}
	\caption{The three different models of activation energy}
	\label{figEmodels}
\end{figure}

The second model is the so-called \emph{kinetic Ising model} \cite{2007NIMPRDjurabekova, 2011PRBReina} (or \emph{final-initial system energy}, as is referred to by Vincent et al.\ \cite{2008JoNMVincenta}). In this model, the activation energy is dependent on the energy difference of the system $\Delta {\cal H}_{ij}$ between the initial $i$ and final states $j$, as well as a migration energy $E_m$, which is a constant determined by the type of defect-atom exchange. Two different forms of activation energy are proposed within this model. The first form is given by \cite{2011PRBReina}:
\begin{equation}
	\Delta E_{ij}= \Bigg\{
	\matrix{
		E_m+\Delta {\cal H}_{ij}, & \text{if}~ \Delta {\cal H}_{ij} > 0 \cr
		E_m,~\quad\quad\quad                   & \text{if}~ \Delta {\cal H}_{ij} < 0}
\end{equation}
This form assumes that the energy barrier of transitions from higher to lower energy states is the migration energy $E_m$, and $E_m+\Delta {\cal H}_{ij}$ otherwise. An alternative, which is used in this work, is given by \cite{2008JoNMVincenta, 2007NIMPRDjurabekova}:
\begin{equation}
	\Delta E_{ij}= E_m+\frac{\Delta{\cal H}_{ij}}{2}
\end{equation}
In this case, the migration energy is considered to be the energy difference between the saddle point and the average energy between states $i$ and $j$, $E_m= E^{SP}-({\cal H}_i+{\cal H}_j)/2$. This definition of $E_m$ results in an expression for $\Delta E_{ij}$ that does not depend of the final state energy ${\cal H}_j$. A schematic diagram showing the different activation energy models discussed here is provided in Fig.\ \ref{figEmodels}. It can be shown that all the three activation energy models satisfy the detailed balance condition, {\it i.e.}:
\begin{equation} \label{DTbal_condition}
	\frac{r_{ij}}{r_{ji}}= \exp \left(-\frac{\Delta {\cal H}_{ij}}{k_BT}\right)
\end{equation}
The different characteristics of each of these models have been discussed in detail by Soisson et al. \cite{2010JoNMSoisson}. In the saddle-point energy model, the height of the energy barrier is not dependent on the energy of the final state, which agrees with the theory of thermally-activated processes. Also, the energy barrier dependence on configurations can be fitted directly from empirical potentials or {\it ab initio} calculations. For its part, the kinetic Ising model assumes that the migration energy depends on the average of the energy difference between the initial and final states. This approach links the energy barrier to the local chemical environment, with the advantage that no knowledge of the saddle-point energy is required. It is also possible to evaluate energy barrier of events other than defect jumps such as recombination and surface reactions (defect annihilation and vacancy creation), described below in Sec. \ref{secEvents}.

\subsubsection{Computing bond energies from electronic-structure calculations.} \label{sec_bondE}
Bond energies to parameterize eq.\ \eref{hclass} and its associated constants $C_{mn}$ can be calculated using a suitable atomistic force fields such as semi-empirical potentials, density-functional theory (DFT), etc. Considering $2^{\text{nd}}$-$nn$ interactions, the following parameters can be used to write a set of equations from which to calculate the bond energies:
\begin{itemize}
\item The cohesive energy of the pure metal A or B can be written as:
\begin{eqnarray}
\label{eq_cohA}
	E_\text{A}^\alpha= -\frac{z_1}{2}\epsilon_{\text{A-A}}^{(1)}-\frac{z_2}{2}\epsilon_{\text{A-A}}^{(2)}\\
	E_\text{B}^\alpha= -\frac{z_1}{2}\epsilon_{\text{B-B}}^{(1)}-\frac{z_2}{2}\epsilon_{\text{B-B}}^{(2)}
\end{eqnarray}
where $z_1$ and $z_2$ are coordination numbers of the first and second nearest neighbor shells, and the superindex $^{(i)}$ refers to the $nn$ shell. Care must be exercised when computing each cohesive energy to ensure that the crystal lattice corresponds to the equilibrium crystal lattice at the desired temperature. 
\item The pair interactions between an A atom and a B atom $\epsilon_{\text{A-B}}$ can be obtained from  the enthalpy of mixing:
\begin{equation}
	E_{mix}=  -\frac{z_1}{2}\left( \epsilon_{\text{A-A}}^{(1)}+\epsilon_{\text{B-B}}^{(1)}-2\epsilon_{\text{A-B}}^{(1)} \right)
	-\frac{z_2}{2}\left( \epsilon_{\text{A-A}}^{(2)}+\epsilon_{\text{B-B}}^{(2)}-2\epsilon_{\text{A-B}}^{(2)} \right)
\end{equation}
\item The formation energy of vacancy is calculated by removing an atom from a perfect lattice position and placing it at the physical limits of the system. For a vacancy in a perfect A-atom matrix containing $N$ lattice sites:
\begin{equation}
	E_{f}^\text{V}= N E_{coh}^\text{A} - (N-1)E_{coh}^\text{A} -z_1\epsilon_{\text{A-V}}^{(1)}-z_2\epsilon_{\text{A-V}}^{(2)}
\end{equation}
\item Similarly, the formation energy of an interstitial pair in an A-atom matrix can be written as:
\begin{equation}
	E_{f}^\text{I}= E_{coh}^\text{A }-z_1\epsilon_\text{A-I}^{(1)}-z_2\epsilon_\text{A-I}^{(2)}
\end{equation}
where I= AA, AB, BB.
\end{itemize}

\subsubsection{Events.} \label{secEvents}
In kMC the kinetic evolution is determined by a series of independent events that represent state transitions. Within the ABVI model, we consider five distinct types of events mediated by point defect mechanisms, discussed below.
\begin{enumerate}
\item \emph{Defect jumps}: vacancies move by exchanging positions with one of the $z_1$ $1^\text{st}$ $nn$ atoms:
$$\text{V} + a\rightarrow a + \text{V}$$
where $a$=A, B. Interstitials, for their part, move via the interstitialcy mechanism introduced above. They can adopt either the \emph{dumbbell} or \emph{crowdion} structure, {\it i.e.}~two atoms sharing one lattice site:
$$\text{I($a_1$-$a_2$)} + a_1 \rightarrow a_1 + \text{I($a_2$-$a_1$)}$$
where an interstitial composed of two atoms $a_1$ and $a_2$ ($a_1,a_2$=A,B) jumps into a neighboring lattice site occupied by atom $a_1$, giving rise to a new interstitial composed of  atoms $a_2$ and $a_1$.

\item \textit{Recombination:} when a vacancy and an interstitial are found within a distance less than a critical distance $r_{c}$, a recombination event occurs. The generic reaction is:
$$\text{I($a_1$-$a_2$)} + V \rightarrow a_1 + a_2$$
Recombinations events occur spontaneously, with $\delta t=0$.

\item \emph{Annihilation at defect sinks}: in this work two types of defect sinks are used. The first one, as suggested by Soisson \cite{2006JoNMSoisson}, is a thin slab of the simulation box designed to act as a perfect defect sink (a simple model of grain boundary). When a defect jumps into a lattice position belonging to the slab, it instantly disappears. To preserve the alloy composition, a `reservoir' is used such that when a vacancy is absorbed at the sink, an atom is randomly chosen from the reservoir and placed at the sink site; for interstitials, one of the two atoms is randomly chosen and stored in the reservoir; the other atom remains on the sink site. 
Another inexhaustible sink is a free surface. The lattice beyond the free surface is considered to be part of a `vacuum' such that atoms adjacent to vacuum lattice sites are defined as `surface atoms'. When a vacancy jumps onto a site occupied by a surface atom, it first switches its position with the atom, and then the vacancy becomes a vacuum site:
$$	\text{V} + a_s \rightarrow a_s + v$$
where $a_s$ refers to a surface atom, and $v$ is a vacuum site. The mechanism for interstitial annihilation is more complex. When an interstitial jumps onto a surface atom site, an instantaneous recombination between the interstitial and the vacuum site occurs (vacuum sites are a special class of vacancies). The reaction can be described as:
$$	\text{I($a_1$-$a_2$)} + v \rightarrow a_1 + a_2 $$

\item \emph{Thermal vacancy emission}: material inhomogeneities such as surfaces, grain boundaries, dislocations, etc, can act as thermal sources of defects. Due to the relatively high energy of interstitial defects compared to vacancies, interstitial emission is often considered negligible. A thermal emission can be regarded as the inverse of a vacancy annihilation event. For a free surface, a vacancy is created just below the surface by having a vacuum site exchange positions with a surface atom:
$$	v + a_s \rightarrow a_s + \text{V} $$
The rate of vacancy emission can become sizable at high temperature, and should not be discarded as an efficient vacancy generation mechanism with a strong effect on the system kinetics.  

\item \emph{Frenkel pair generation}: when considering irradiation with light particles (e.g., electrons), V-I pairs are generated in the lattice. As implemented in our method, when a Frenkel pair insertion occurs, two lattice sites are randomly chosen, one becomes a vacancy and the other becomes an interstitial formed by the two atoms involved:
$$a_1 + a_2 \rightarrow \text{V + I($a_1$-$a_2$)}$$
Frenkel pairs are introduced at a rate consistent with the imposed irradiation dose rate (usually measured in displacements per atom per second, or dpa$\cdot$s$^{-1}$).
\end{enumerate}
A compilation of all the reactions and events discussed in this section is provided in Table \ref{table_rules}.
\begin{table}[h]
	\caption{Event reactions considered in this work. V: vacancy, A: matrix atom, B: solute atom, AA: self interstitial, AB: mixed interstitial, BB: pure solute interstitial, $v$: vacuum atom, A$_s$: surface matrix atom, B$_s$: surface solute atom.}
	\label{table_rules}
	\centering
	\begin{tabular}{| c | c | c | c |}
	\hline
	Vacancy jumps & Interstitial jumps & Recombinations & Frenkel pair generation \\
	\hline
	V+A$\rightarrow$A+V & AA+A$\rightarrow$A+AA & AA+V$\rightarrow$A+A & A+A$\rightarrow$AA+V \\
	V+B$\rightarrow$A+B & AA+B$\rightarrow$B+AA & AB+V$\rightarrow$A+B & A+B$\rightarrow$AB+V \\
	& BB+A$\rightarrow$B+AB & BB+V$\rightarrow$B+B & B+B$\rightarrow$BB+V\\
	& BB+B$\rightarrow$B+BB & & \\
	& AB+A$\rightarrow\Bigg\{\matrix{\text{A+AB}\cr\text{B+AA}}$ & & \\
	& AB+B$\rightarrow\Bigg\{\matrix{\text{A+BB}\cr\text{B+AB}}$ & & \\
	\hline
	\end{tabular}
	\begin{tabular}{| c | c | c |}
	\hline
	\multicolumn{2}{|c|}{Defect annihilation} & \multirow{2}{*}{Thermal emission} \\
	\cline{1-2}
	Ideal sink & Surface & \\
	\hline
	V$\rightarrow$A & V+A$_s$$\rightarrow$A$_s$+$v$ & $v$+A$_s$$\rightarrow$A$_s$+V \\
	V$\rightarrow$B & V+B$_s$$\rightarrow$B$_s$+$v$ & $v$+B$_s$$\rightarrow$B$_s$+V \\
	AA$\rightarrow$A & AA+$v$$\rightarrow$A+A$_s$ & \\
	BB$\rightarrow$B & BB+$v$$\rightarrow$B+B$_s$ & \\
	AB$\rightarrow\Bigg\{\matrix{A\cr B}$& AB+$v$$\rightarrow\Bigg\{\matrix{\text{A+B$_s$}\cr\text{B+A$_s$}}$ & \\
	\hline
	\end{tabular}
\end{table}

\section{Results} \label{sec_results}
This section consists of various verification checks undertaken to ensure the correctness of our approach.
The first tests are designed to check the `downward' consistency of our model, {\it i.e.}~comparing against AV and ABV models with reduced complexity w.r.t.~the ABVI Hamiltonian\footnote{The AV case --as studied by Reina et al.\ \cite{2011PRBReina}-- was trivially reproduced by our method, and for brevity we omit any further discussion on it.}. Subsequently, we compare our method with KMC simulations of three different ABVI systems published in the literature. In all simulations, atoms are initially assigned randomly to lattice sites so as to achieve a perfect solid solution as a starting configuration.

\subsection{ABV system: Precipitation of Fe-Cu alloys}

\begin{figure}[h]
	\centering
	\begin{subfigure}{0.45\textwidth}
		\centering
		\includegraphics[width=\textwidth]{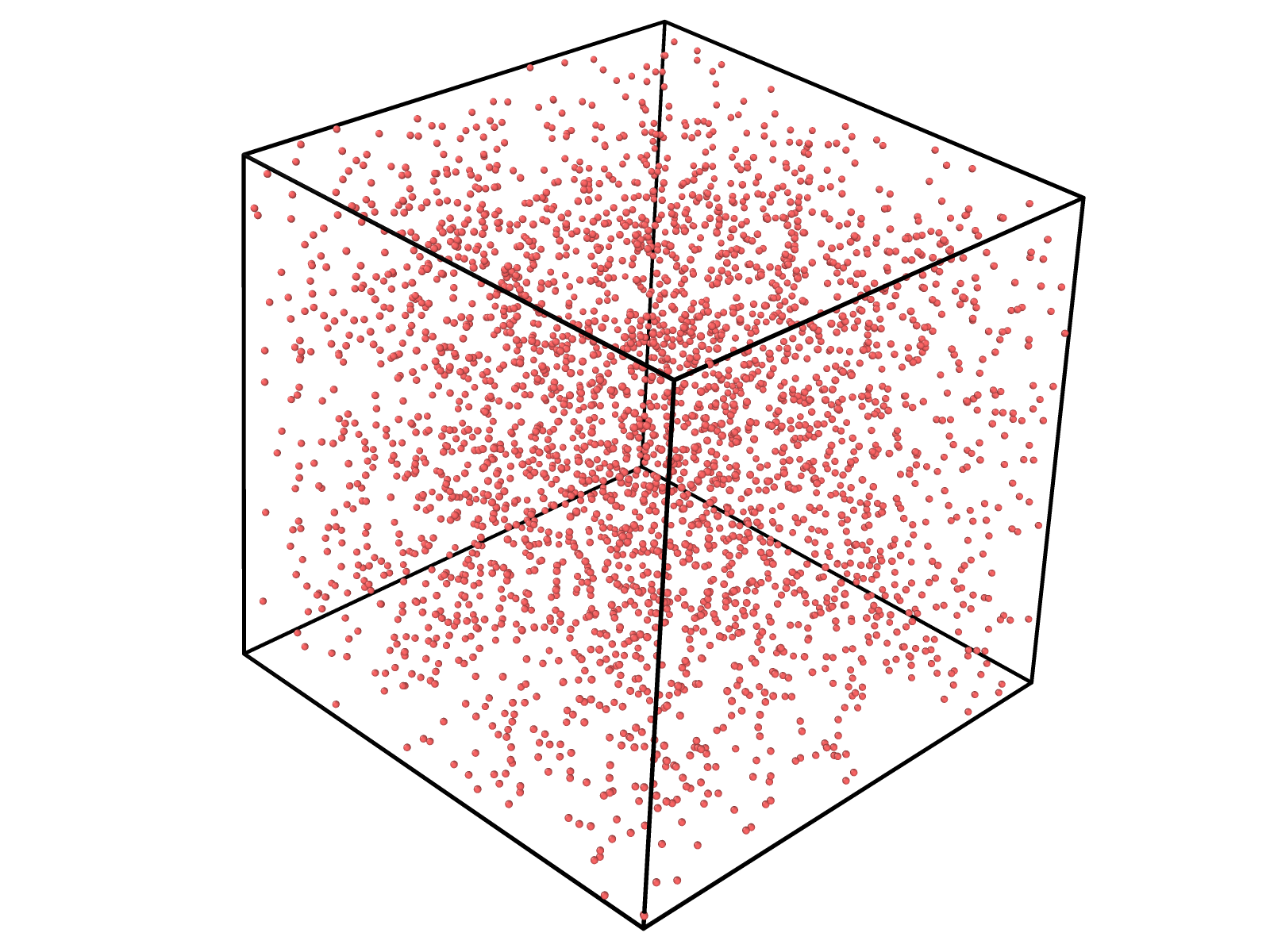}
		\caption{}
	\end{subfigure}
	\begin{subfigure}{0.45\textwidth}
		\centering
		\includegraphics[width=\textwidth]{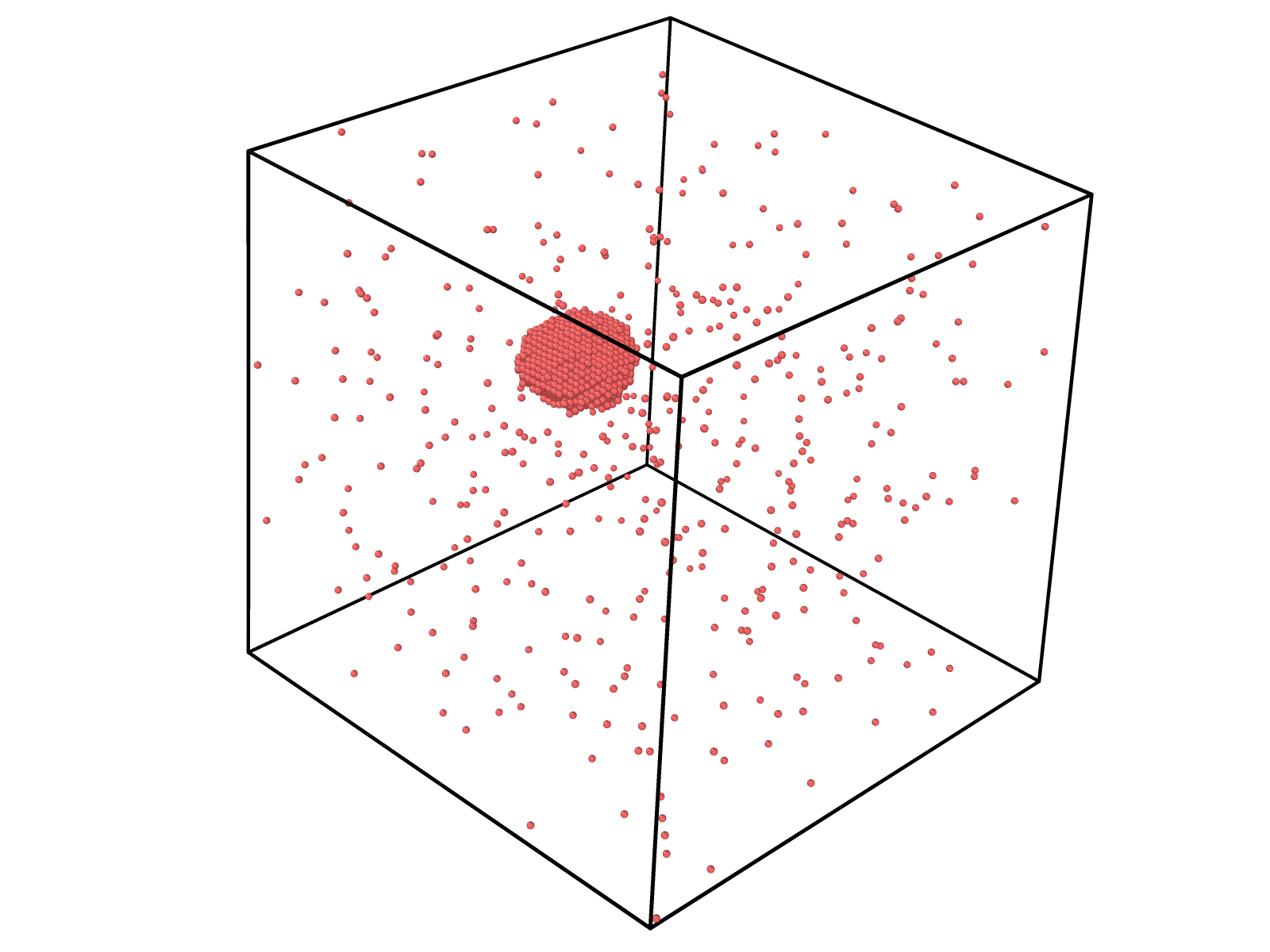}
		\caption{}
	\end{subfigure}
	\caption{Initial (a), $t=0$, and final (b), $t=28368$ s alloy configurations. The red dots represent solute atoms (B atoms). Solvent atoms and the vacancy are omitted for clarity.}
	\label{figABVsnap}
\end{figure}

\begin{figure}[tbh]
	\centering
	\includegraphics[width=0.5\textwidth]{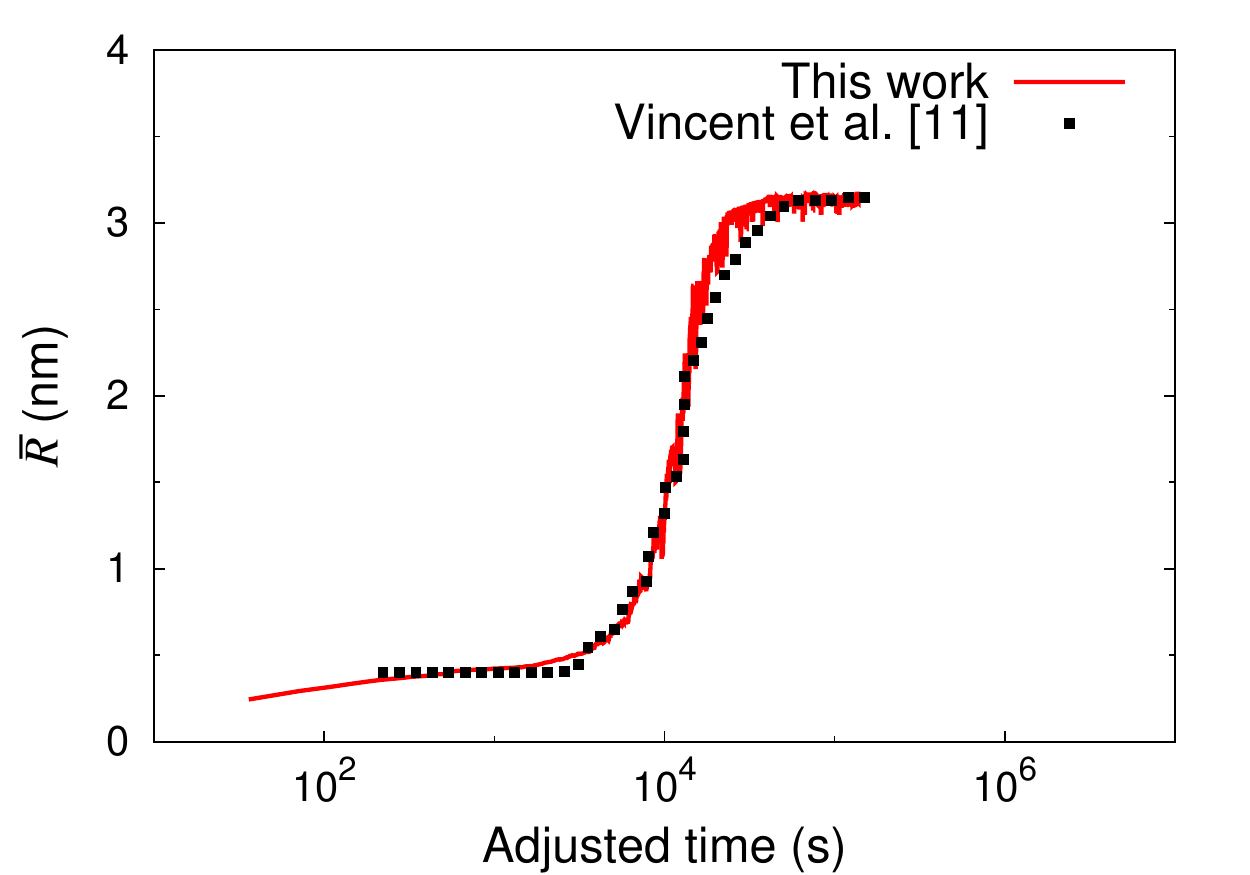}
	\caption{The cluster mean radius of the ABV Fe-Cu system. The red line represents the results in this work; the black filled squares are the data from Vincent et al. \cite{2008JoNMVincenta}}
	\label{figABVr}
\end{figure}

First we simulate the system considered by Vincent et al.~\cite{2008JoNMVincenta}: a Fe-0.6\% at.~Cu alloy occupying a periodic BCC lattice arranged into computational box with $80\times80\times80$ primitive cells containing 512,000 atoms and a single vacancy. The Hamiltonian includes $2^\text{nd}$-$nn$ interactions with energy coefficients given in Table \ref{table_coeff1}. The energies of mixing for $1^\text{st}$ and $2^\text{nd}$-$nn$ are 0.26 and 0.24 eV, which suggest a strong tendency toward phase separation \cite{2008TMGaskell}. The temperature is fixed at 773 K. During the simulations, the vacancy may become trapped in solute precipitates, which does not result in net microstructural evolution and may stall the simulations. To correct for this, Vincent et al.~proposed to increment the kMC time only when the vacancy is surrounded by at most one solute atom. As well, to account for an unrealistically high vacancy concentration, the kMC time step was rescaled according to:
\begin{equation}
\delta t= \frac{C_\text{V}^0}{C_\text{V}^\text{kMC}}\delta t_\text{kMC}
\end{equation}
where $C_\text{V}^0=\exp\left(-E^\text{V}_f/k_BT\right)$ is the thermodynamic vacancy concentration. However, Vincent et al.~adjust their kMC time by comparing the kinetic evolution directly with experiments. By way of example, they matched a cluster mean radius of 0.9 nm in their to a time of 7200 s. For consistency, we adopt the same approach here.							
\begin{table}[h]
	\caption{Bond energies for the Fe-Cu ABV system. A represents Fe atoms, B Cu atoms, and V is the vacancy.}
	\label{table_coeff1}
	\centering
	\begin{tabular}{| c | c | c | c | c | c | c |}
		\hline
		\multicolumn{5}{|c|}{$1^\text{st}$-$nn$ interactions (eV)} &
		\multicolumn{2}{|c|}{Migration energy (eV)} \\
		\hline
		$\epsilon_{\text{A-A}}^{(1)}$ & 
		$\epsilon_{\text{A-B}}^{(1)}$ & 
		$\epsilon_{\text{B-B}}^{(1)}$ &
		$\epsilon_{\text{A-V}}^{(1)}$ & 
		$\epsilon_{\text{B-V}}^{(1)}$ &
		$\ \ \ E_m^{V-A}\ \ \ $ & 
		$E_m^{V-B}$ \\
		\hline
		$-0.611$ & $-0.480$ & $-0.414$ & $-0.163$ & $-0.102$ & $0.62$ & $0.54$ \\
		\hline
		\multicolumn{5}{|c|}{$2^\text{nd}$-$nn$ interactions (eV)} &
		\multicolumn{2}{|c|}{Jump frequency (s$^{-1}$)} \\
		\hline
		$\epsilon_{\text{A-A}}^{(2)}$ & 
		$\epsilon_{\text{A-B}}^{(2)}$ & 
		$\epsilon_{\text{B-B}}^{(2)}$ &
		$\epsilon_{\text{A-V}}^{(2)}$ & 
		$\epsilon_{\text{B-V}}^{(2)}$ &
		$\nu_\text{A}^\text{V}$ &
		$\nu_\text{B}^\text{V}$ \\
		\hline
		$-0.611$ & $-0.571$ & $-0.611$ & $-0.163$ & $-0.180$ &
		$6\times 10^{12}$ & $6\times 10^{12}$ \\ 
		\hline
	\end{tabular}
\end{table}							
The initial and final configurations are shown in Fig. \ref{figABVsnap}. 
The kinetic evolution of precipitation is quantified by calculating the cluster mean radius of solute atoms as a function of time. It is assumed that a B atom belongs to a cluster if one of its $1^\text{st}$-$nn$ is also a B atom of the cluster. The cluster size is computed assuming a spherical shape from the expression \cite{2002CMSSchmauder}:
\begin{equation}
	\bar R=a_0\left(\frac{3N}{8\pi}\right)^{\frac{1}{3}}       
\end{equation}
where $\bar R$ is the cluster mean radius, $N$ is the number of solute atoms in the cluster, and $a_0$ is the lattice constant of the BCC lattice. As in ref.\ \cite{2008JoNMVincenta}, clusters containing three or less atoms are not counted towards the calculation of $\bar R$. Figure \ref{figABVr} shows our data compared to those of Vincent et al. At about $10^4$ s, a transition in $\bar R$ occurs, where the cluster size grows abruptly before leveling off at longer times. Although our model captures the timescale evolution of $\bar R$, a factor of 1.6 was found among our data and theirs. The source of this discrepancy is unknown, although we attribute it to artifacts related to fits of the experimental data.

\subsection{ABVI system: Solute segregation at sinks} \label{secABVI1}
In this test, we reproduce the work of Soisson \cite{2006JoNMSoisson}. The system consists of a BCC $256\times64\times64$ triclinic crystal lattice containing an A-5\%B alloy, vacancies and interstitials defects. A perfect planar defect sink is placed in the middle of the crystal and kMC simulations of (radiation-induced) segregation at the defect sink are performed. Frenkel pairs are generated at a rate of $G= 10^{-6}$ dpa$\cdot$s$^{-1}$ following the mechanism described in Sec.\ \ref{secEvents}. 

\begin{figure}[h]
	\centering
	\begin{subfigure}{0.45\textwidth}
		\centering
		\includegraphics[width=\textwidth]{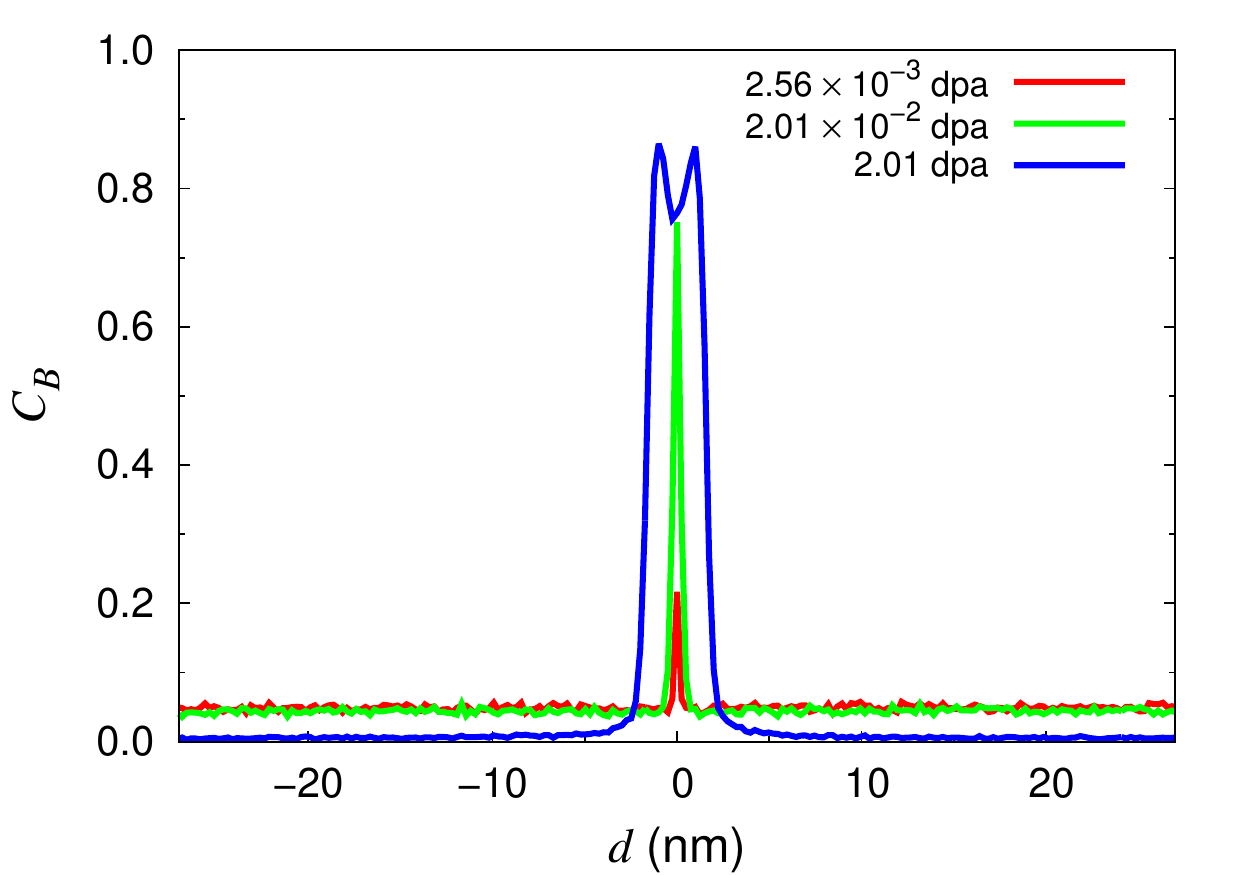}
		\caption{}
	\end{subfigure}
	\begin{subfigure}{0.45\textwidth}
		\centering
		\includegraphics[width=\textwidth]{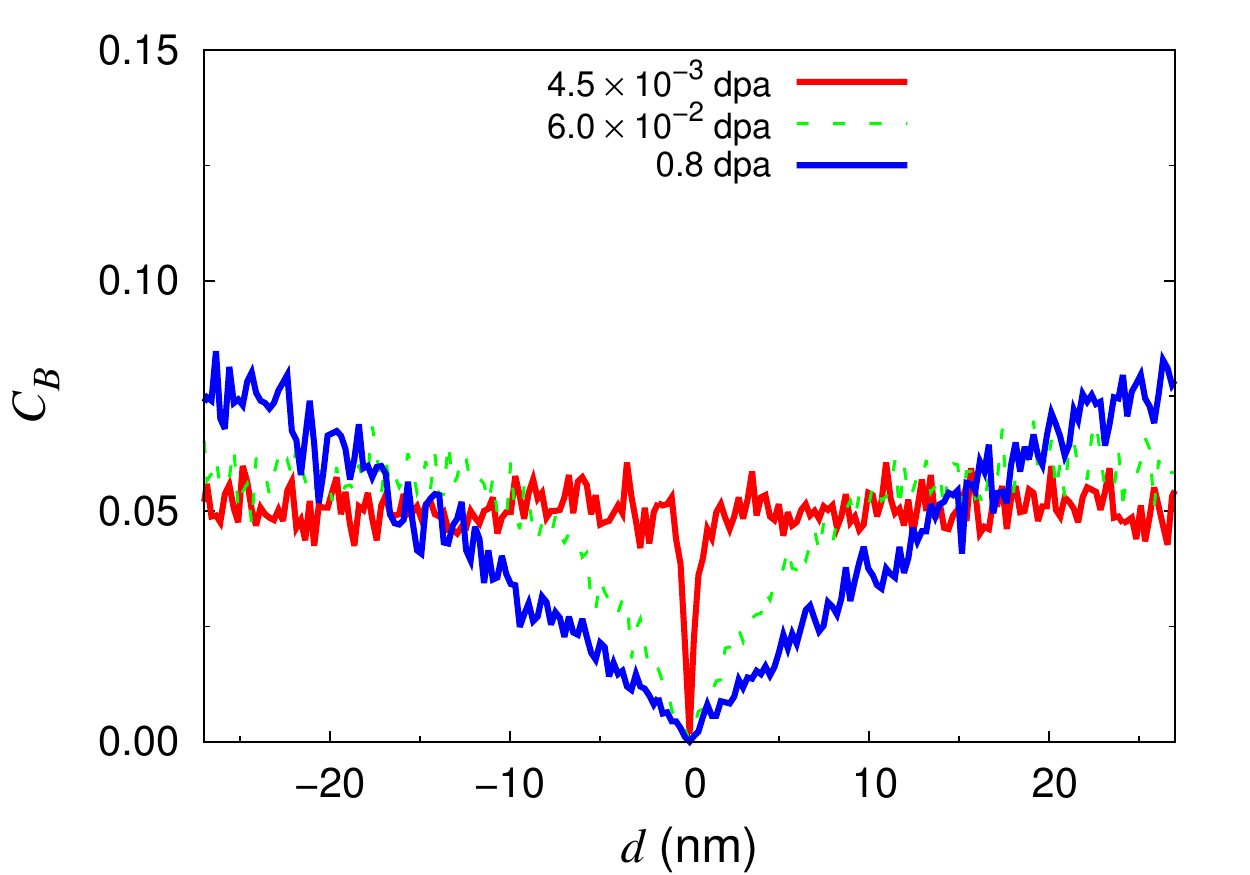}
		\caption{}
	\end{subfigure}
	\begin{subfigure}{0.45\textwidth}
		\centering
		\includegraphics[width=\textwidth]{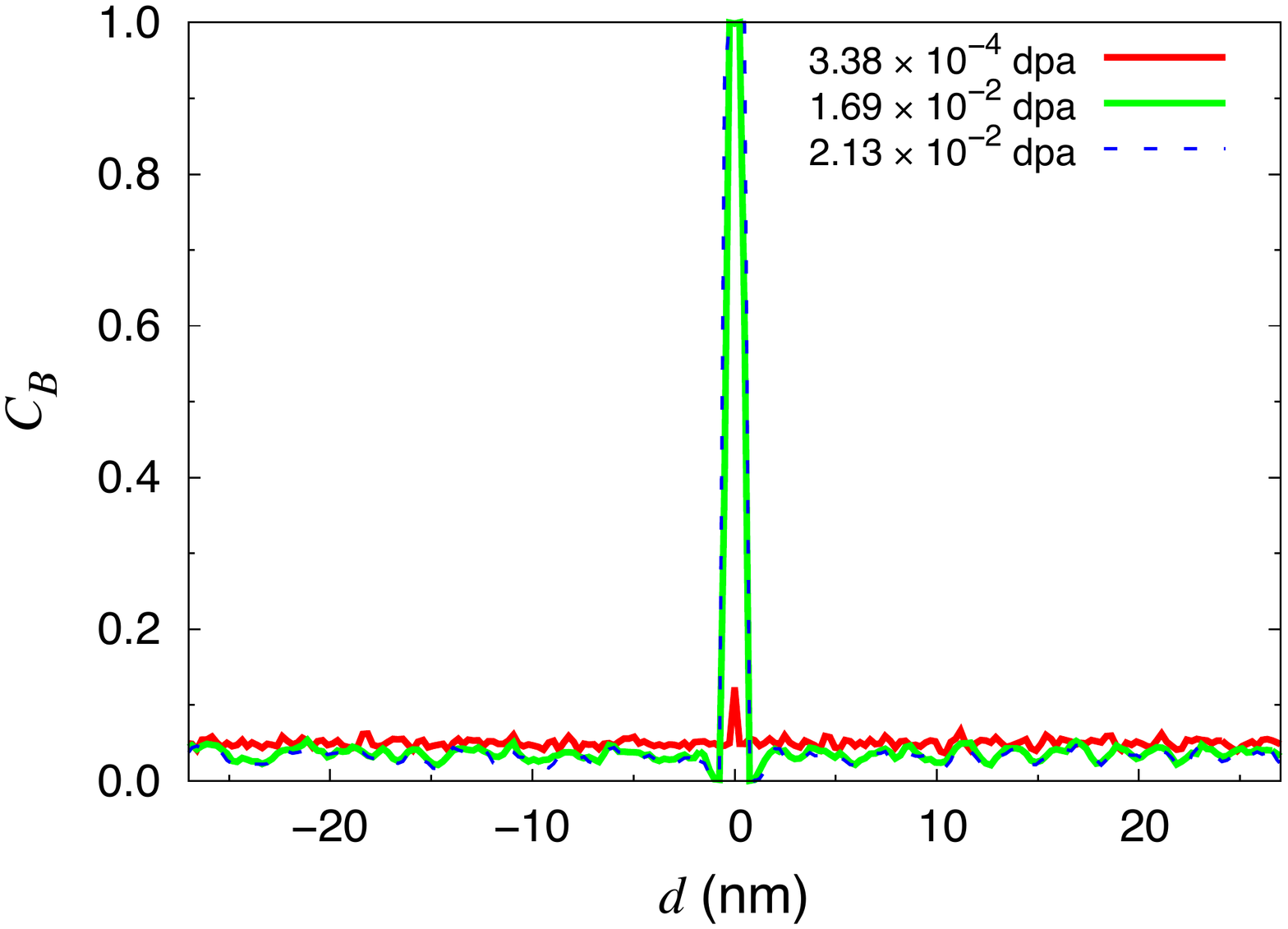}
		\caption{}
	\end{subfigure}
	\begin{subfigure}{0.45\textwidth}
		\centering
		\includegraphics[width=\textwidth]{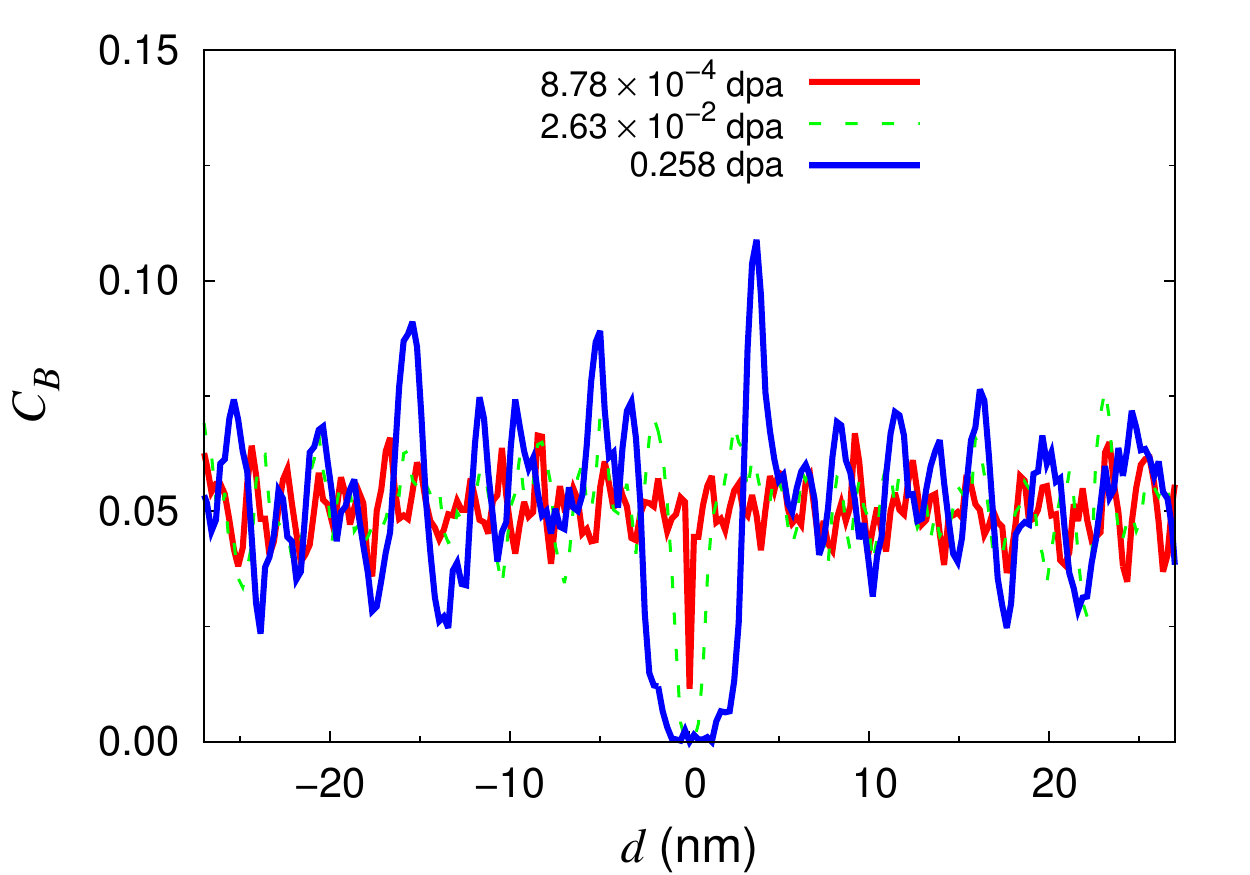}
		\caption{}
	\end{subfigure}
	\caption{Spatial solute concentration profiles at different doses for the undersatured alloy for the (a) solute enrichment and (b) solute depletion cases at $T= 800$ K. The supersatured case for (c) solute enrichment and (d) solute depletion at $T= 500$ K are also shown. The nominal solute concentration of the alloy is $C_B= 0.05$ and the dose rate is $10^{-6}$ dpa$\cdot$s$^{-1}$.}
	\label{figABVIcb}
\end{figure}

Segregation at the sinks is governed by the onset of solute fluxes in the system. These fluxes are mediated by defect migration to and absorption at the sink. The solute flux can be controlled by setting the defect migration energies such that exchanges with B atoms are preferred over exchanges with A atoms (or vice versa), resulting in enrichment or depletion of solute at the defect sink.
While Soisson uses a saddle-point model to obtain the activation energy (cf.\ Sec.\ \ref{sec_actE}), our implementation of the ABVI Hamiltonian has been design to employ a kinetic Ising model. In order to make both approaches as close to one another as possible, we use Soisson's bond energies directly and adjust the migration energies $E_m$ so as to match the kinetic evolution. The parameters used are shown in Table \ref{table_coeff2}. There are four sets of parameters. The first two, ABVI-1 and ABVI-2, correspond to a system with relatively low energy of mixing ($E_{mix}= 0.216$ eV), representing undersaturated solid solutions with high solubility limits. The other two, ABVI-3 and ABVI-4 correspond to a system with $E_{mix}=0.680$ eV leading to supersaturated solid solutions. 
Systems ABVI-1 and ABVI-3 $E_m$ are such that a net flux of B atoms develops toward the sink ($E_m^\text{V-A}<E_m^\text{V-B}$; $E_m^\text{I-A}>E_m^\text{I-B}$), whereas ABVI-2 and ABVI-4 result in solute depletion at the sink --the so-called \emph{inverse Kirkendall} effect--($E_m^\text{V-A}>E_m^\text{V-B}$; $E_m^\text{I-A}<E_m^\text{I-B}$). For simplicity, migration energies of vacancies and interstitials are set to produce the same segregation tendency for each set of parameters.
\begin{table}[h]
	\caption{Parameters for the ABVI system (after Soisson \cite{2006JoNMSoisson}). `A' and `B'  denote solvent and solute atoms, respectively. `V' represents vacancies and `I' all types of interstitial defects. All energies given in eV. Attempt frequencies given in Hz.}
	\label{table_coeff2}
	\centering
	\begin{tabular}{| c | c | c | c | c |}
		\hline 
		\multirow{3}{*}{Kinetic parameters} & ABVI-1 & ABVI-2 & ABVI-3 & ABVI-4 \\
		\cline{2-5}
		& \multicolumn{2}{|c|}{high solubility} 
		& \multicolumn{2}{|c|}{low solubility} \\
		\cline{2-5}
		& enrichment & depletion & enrichment & depletion \\
		\hline
		$\nu_\text{A}^\text{V}=\nu_\text{B}^\text{V}=\nu_\text{A}^\text{I}=\nu_\text{B}^\text{I}$ &
		$5\times10^{15}$ & $5\times10^{15}$ & 
		$5\times10^{15}$ & $5\times10^{15}$ \\
		\hline
		$\epsilon_{\text{A-A}}=\epsilon_{\text{B-B}}$  & 
		$-1.07$ & $-1.07$ & $-1.07$ & $-1.07$ \\
		\hline
		$\epsilon_{\text{A-B}}$ \   &
		$-1.043$ & $-1.043$ & $-0.985$ & $-0.985$ \\
		\hline
		$\epsilon_{\text{A-V}}=\epsilon_{\text{B-V}}$   &
		$-0.3$ & $-0.3$ & $-0.3$ & $-0.3$ \\
		\hline
		$\epsilon_{A\text{-}I}=\epsilon_{B\text{-}I}$   &
		0 & 0 & 0 & 0 \\
		\hline
		$E_m^{V-A}$  & 
		0.95 & 1.1 & 0.8 & 1.05 \\
		\hline
		$E_m^\text{V-B}$  & 
		1.05 & 0.9 & 1.2 & 0.95 \\
		\hline
		$E_m^\text{I-A}$  & 
		0.5 & 0.35 & 0.55 & 0.2 \\
		\hline
		$E_m^\text{I-B}$  & 
		0.5 & 0.65 & 0.45 & 0.8 \\
		\hline
	\end{tabular}
\end{table}
Other details considered by Soisson, such as recombination radii, event sampling, etc., are also followed here\footnote{With one exception: the Frenkel pair distance is not set in this work.}.
The spatial solute concentration profiles are shown in Fig.\ \ref{figABVIcb}. 

\begin{figure}[h]
	\centering
	\begin{subfigure}{0.3\textwidth}
		\centering
		\includegraphics[width=\textwidth]{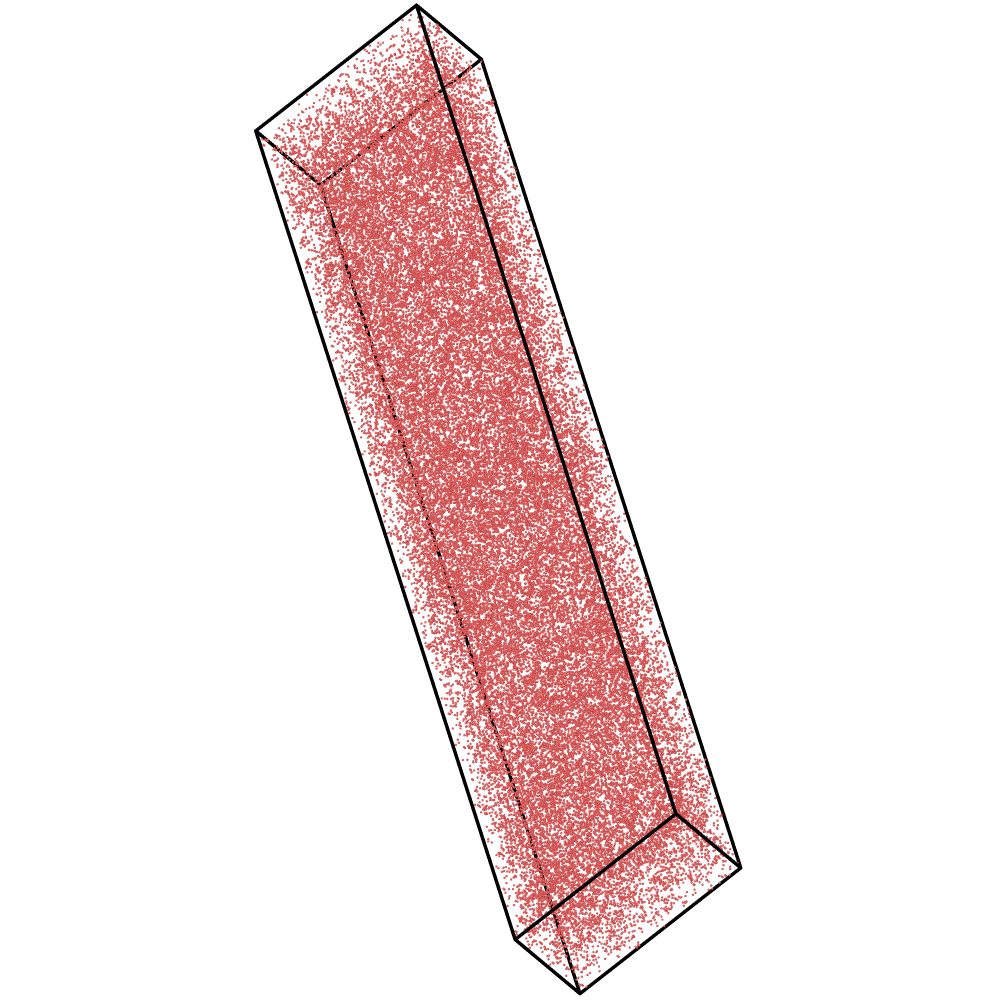}
		\caption{}
	\end{subfigure}
	\begin{subfigure}{0.3\textwidth}
		\centering
		\includegraphics[width=\textwidth]{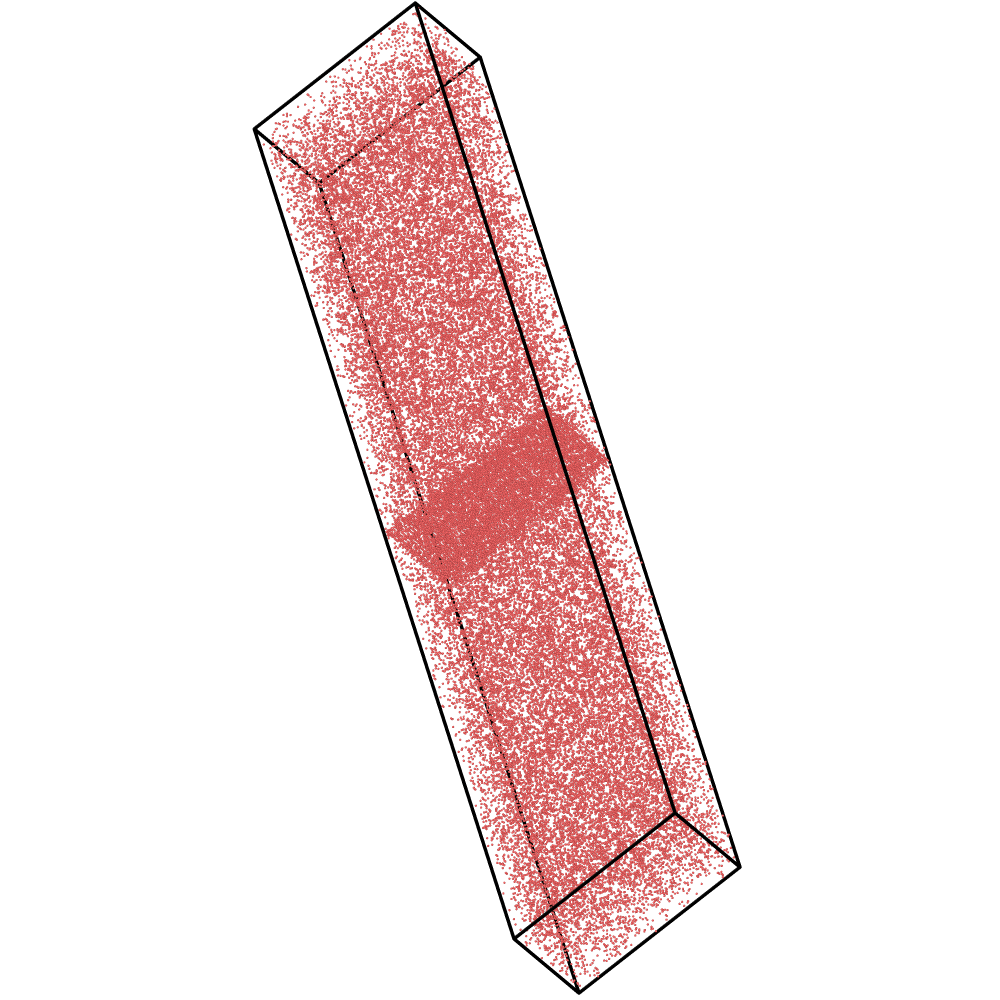}
		\caption{}
	\end{subfigure}
	\begin{subfigure}{0.3\textwidth}
		\centering
		\includegraphics[width=\textwidth]{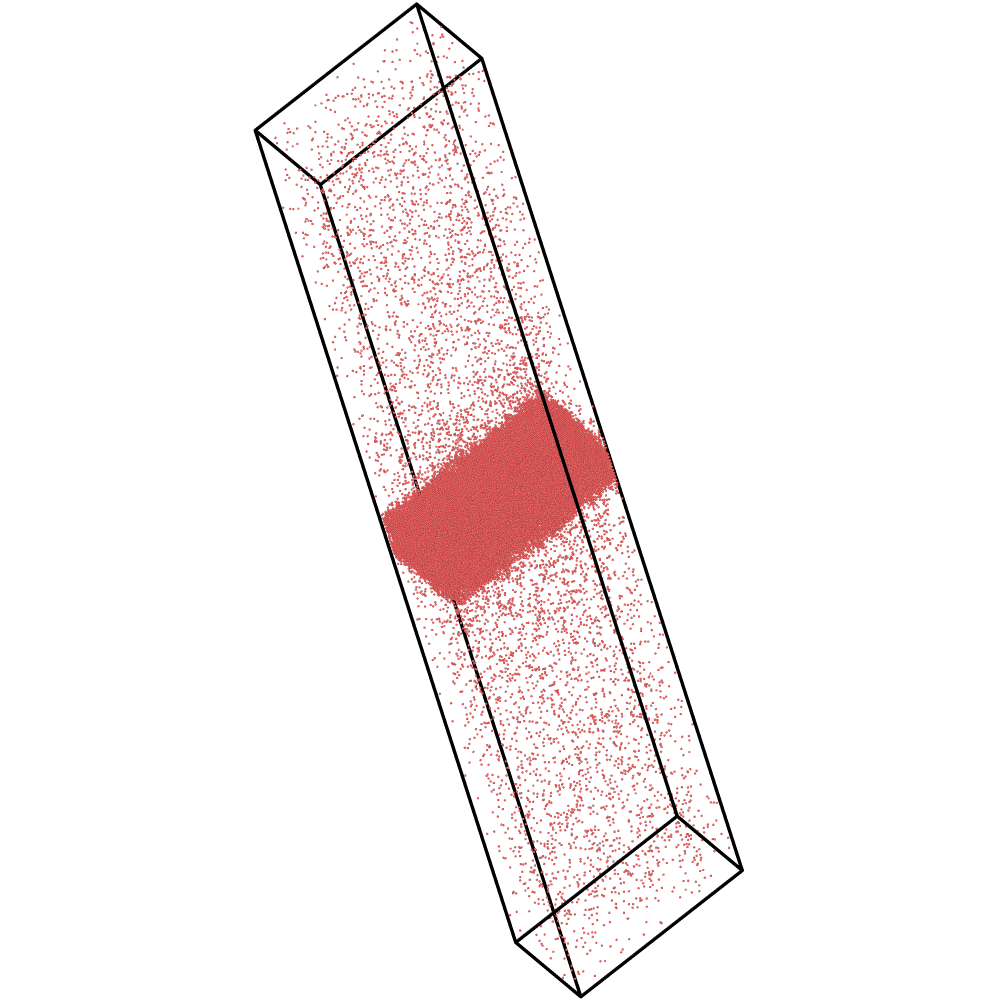}
		\caption{}
	\end{subfigure}
	\begin{subfigure}{0.3\textwidth}
		\centering
		\includegraphics[width=\textwidth]{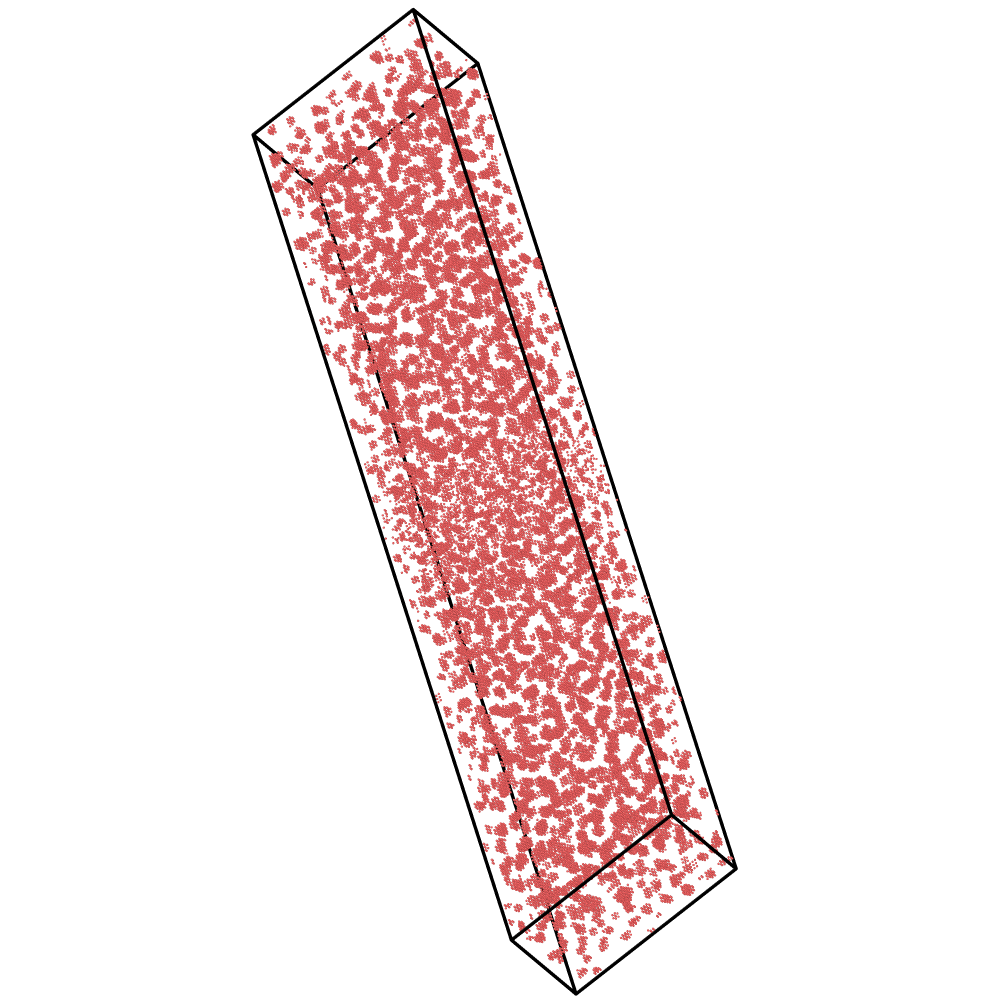}
		\caption{}
	\end{subfigure}
	\begin{subfigure}{0.3\textwidth}
		\centering
		\includegraphics[width=\textwidth]{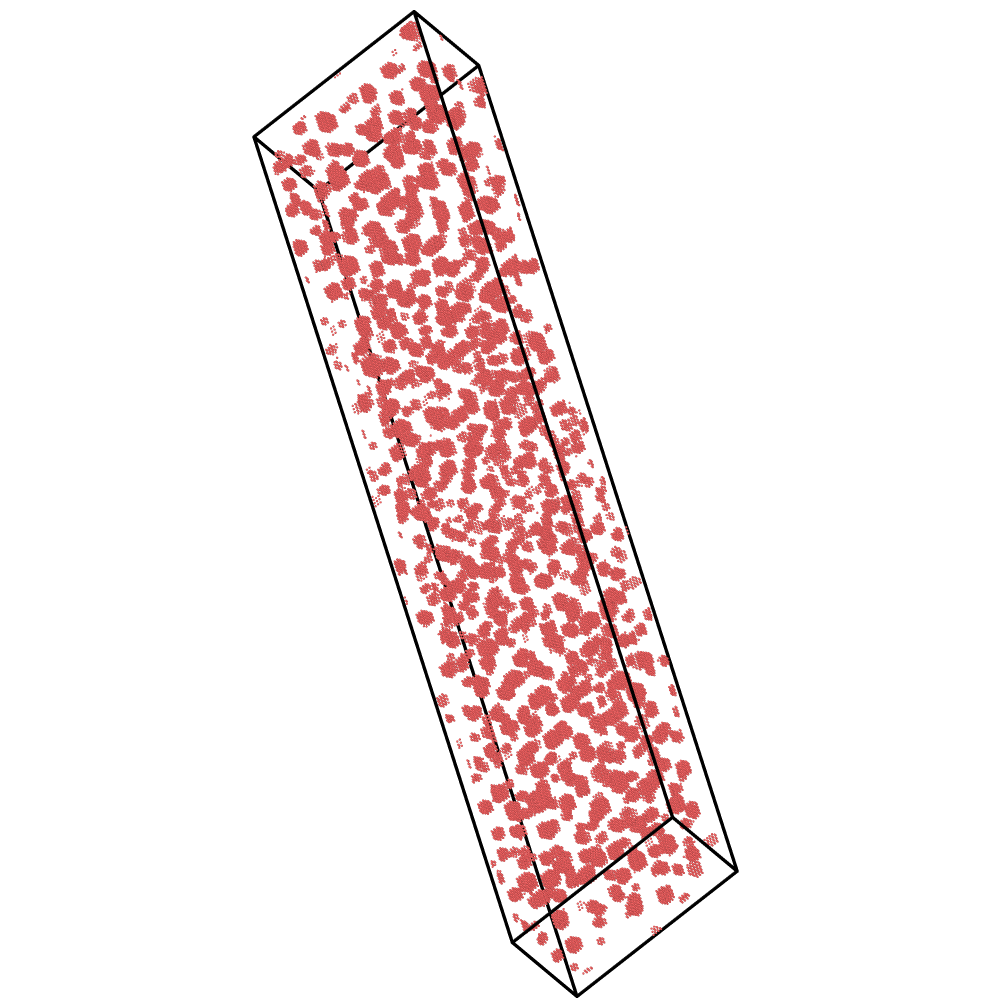}
		\caption{}
	\end{subfigure}
	\begin{subfigure}{0.3\textwidth}
		\centering
		\includegraphics[width=\textwidth]{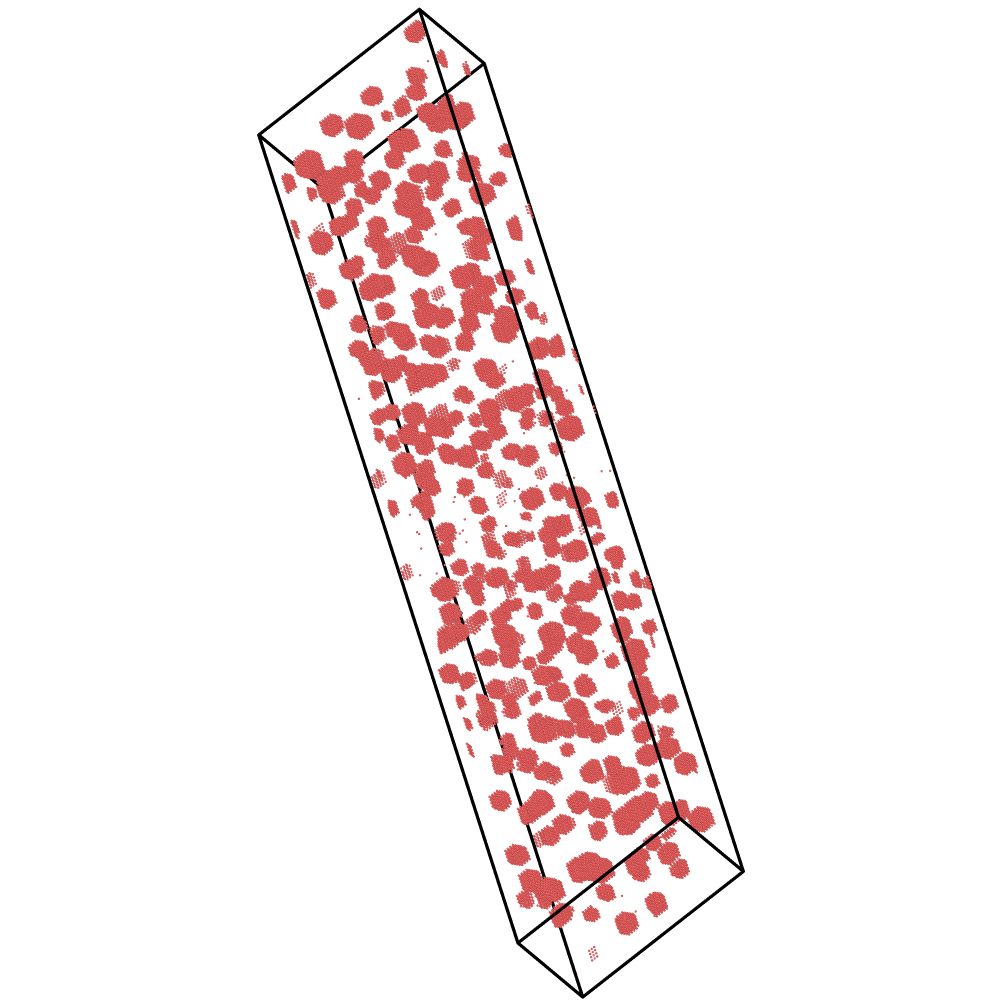}
		\caption{}
	\end{subfigure}
	\caption{Snapshots of ABVI-1 system (undersaturated, enrichment) at (a) $2.56\times10^{-3}$ (b) $2.01\times10^{-2}$ and (c) $2.01$ dpa. For the ABVI-4 case (supersaturated, depletion), configurations are shown at (d) $8.78\times10^{-4}$ (e) $2.63\times10^{-2}$ and (f) $0.258$ dpa. Only solute atoms are shown.}
	\label{fig_ABVIsnap}
\end{figure}

In the undersaturated alloy, no precipitation in the bulk is observed. As the dose increases, the concentration of B atoms near the sink is enhanced (reduced) for the enrichment (depletion) parameter set.
For the enrichment case ABVI-1, a solute concentration drop at the center of the system is observed. This can rationalized in terms of interstitialcy jumps. After the solute concentration raises near the sink, interstitials must traverse a solute-rich region in order to reach the sink. As interstitials penetrate the near-sink region, they will increasingly become of the AB type. Because $\epsilon_\text{A-B} > \epsilon_\text{B-B}$, A atoms located in this solute-rich region are energetically unfavorable. Therefore, interstitials jumps favor the avoidance of A-B bonds, which results in enhanced matrix atom transport to the sink. This phenomenon was not observed in Soisson's work because they used a saddle-point energy model that gives a nonlocal activation energy (does not depend on the atomic environment of the jumping atom). 
Increasing the driving force for solute transport toward the sink (e.g., by setting $E_m^\text{I-A}=0.6,~E_m^\text{I-B}=0.4$), the drop at the sink disappears.  Snapshots for ABVI-1 and ABVI-4 at three different doses are shown in Fig. \ref{fig_ABVIsnap}.

For the low solubility alloy, on the other hand, bulk precipitation does occur, as one would expect given the low marginal difference between bulk and sink segregation driving forces. As Fig.\ \ref{figABVIcb} shows, the solute spatial profiles are much more fluctuative than their high solubility counterparts, especially for the depletion case (ABVI-2 vs.\ ABVI-4). This of course is a manifestation of the formation of precipitates in bulk. The mean free path for solute diffusion is quite low, due to a high number density of precipitates acting as trapping sites, which makes depletion dynamics slow. Soisson observed a less intense bulk precipitation than shown here, possibly also due to the different in activation energy models employed. In any case, the global qualitative features of the alloy evolution kinetics are matched by both methods.

\subsection{ABVI system: Radiation-induced segregation at surfaces}
The last verification example that we tackle in this paper is that of a finite system containing a binary alloy under irradiation. 
This mimics the case considered by Dubey and El-Azab, which studied binary Au-Cu alloy under irradiation using a two-dimensional continuum reaction-diffusion model bounded by a free surface \cite{2015CMSDubey}. These authors use effective rate theory to solve the ordinary differential equation system representing defect kinetics with spatial resolution. As such, our method differs fundamentally in that it relies on a discrete lattice description, and so the comparison between both approaches must account for this distinction. Our lattice system, however, is constructed so as to create two free surfaces along one of the dimensions of the computational cell, with periodic boundary condition used for the other two. Adjacent to the free surfaces, several layers of `vacuum' atoms are introduced (cf.\ Sec.\ \ref{secEvents} for the mechanisms involving these vacuum atoms). In this fashion, the surface is always univocally defined as the interface between atomic lattice sites and vacuum sites, which provides a convenient way to study the surface roughness as simulations progress.

Considering free surfaces introduces both a defect sink and a source. In addition to Frenkel-pair generation by irradiation, point defects can also be emitted thermally from the surface. Following Dubey and El-Azab, Frenkel-pair generation rate is set at $1.0$ dpa$\cdot$s$^{-1}$. 
Regarding defect emission from the surface, the high formation energy difference between interstitial defects and vacancies allows us to discount thermal emission of SIAs, as done in ref.\ \cite{2015CMSDubey}, while vacancies can be created at all surface sites. In each step, the rates of all the possible creation paths, {\it i.e.}~all 1$^\text{st}$-$nn$ jumps from surface sites towards the interior of the box, are calculated and added to the global kMC event list\footnote{Vacancy emission can occur from any surface site. Given the large number of such sites, we precompute all the thermal emission rates at the beginning, and then simply update the list when the local chemical environment around a surface site changes during the kMC simulation.}. 

The annihilation of defects at surfaces is also considered, as described in Sec.~\ref{secEvents}. After Dubey and El-Azab, we study a face-centered cubic binary Au-10\% at.\ Cu alloy using the energetics provided in Table \ref{table_coeff3} based on a study by Hashimoto et al.\ \cite{1995JoNMHashimoto}. The computational box dimensions are $660\times270\times4$ primitive cells, with a vacuum buffer of 20 atomic layers on either side of the free surface, along the $x$-direction. In this case, jumps of mixed interstitials are calculated considering both directional possibilities, e.g.\ AB+A$\rightarrow$B+AA, or AB+A$\rightarrow$AB+A (cf.\ Table \ref{table_rules}), with their total rate weighted by a factor of $\nicefrac{1}{2}$ to preserve the correct sampling statistics. 

In this work, we set the vacuum energy level as the zero reference, {\it i.e.}~$\epsilon_{v\text{-X}}=0$ (where $X$= A, B, V, $v$), and the energies of atoms on the surface are simply tallied in terms of the number of missing surface bonds. The defect bond energy parameters then can be obtained from formation energies of vacancy and interstitial using the formulas described in Sec.\ \ref{sec_bondE}. The surface energy per area and defect formation energies are taken from Dubey and El-Azab's paper. 
In addition, after Hashimoto et al., a conversion energy is applied when interstitial defects change their type after a diffusive jump. On some occasions, the activation energy for interstitialcy jumps can become negative, which we simply interpret as a spontaneous event within the kMC cycle.
\begin{table}[h]
\centering
	\caption{The parameters for the Au-Cu ABVI system. `A' are Cu atoms, `B' are Au atoms. X, Y= A, B; Z= A, B, V, $v$.}
	\label{table_coeff3}
	\begin{tabular}{| c | c | c | c | c |}
		\hline
		\multicolumn{5}{| c |}{Bond energies (eV)} \\
		\hline
		   $\epsilon_{\text{X-Y}}$
		& $\epsilon_{\text{V-X}}$
		& $\epsilon_{\text{AA-X}}$
		& $\epsilon_{\text{AB-X}}$
		& $\epsilon_{\text{BB-X}}$\\
		\hline
		$-0.1425$ & $-0.01625$ & 0.24625 & 0.12875 & 0.14625 \\
		\hline
		\multicolumn{5}{| c |}{Migration energies (eV)} \\
		\hline
		   $E_m^\text{V-A}$
		& $E_m^\text{V-B}$
		& $E_m^\text{I-AA}$
		& $E_m^\text{I-AB}$
		& $E_m^\text{I-BB}$ \\
		\hline
		0.88 & 0.76 & 0.3 & 0.377 & 0.12 \\
		\hline
		\multicolumn{5}{| c |}{Conversion energies (eV)} \\
		\hline
		  $E_c^{\text{AA}\rightarrow \text{AB}}$
		& $E_c^{\text{AB}\rightarrow \text{AA}}$
		& $E_c^{\text{BB}\rightarrow \text{AB}}$
		& $E_c^{\text{AB}\rightarrow \text{}BB}$ & \\
		\hline
		0.3 & 0.5 & 0.12 & 0.32 & \\
		\hline				
	\end{tabular}
\end{table}

\begin{figure}[h]
	\centering
	\includegraphics[width=0.7\textwidth]{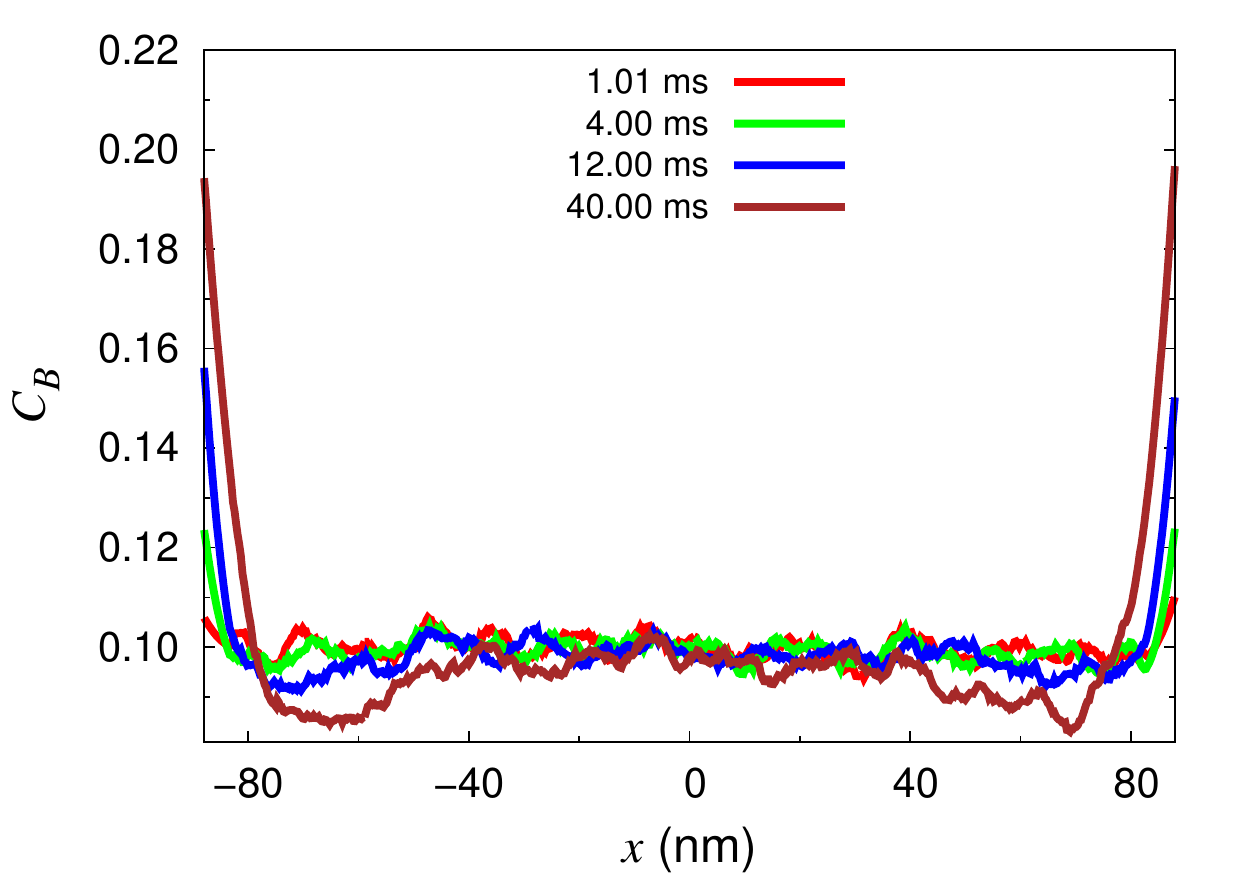}
	\caption{Solute concentration profile for different doses for the Au-10\% at.\ Cu alloy at 650 K . The dose rate is 1.0 dpa$\cdot$s$^{-1}$.}
	\label{figSRFprofile}
\end{figure}

\begin{figure}[tbh]
	\centering
	\includegraphics[width=0.8\textwidth]{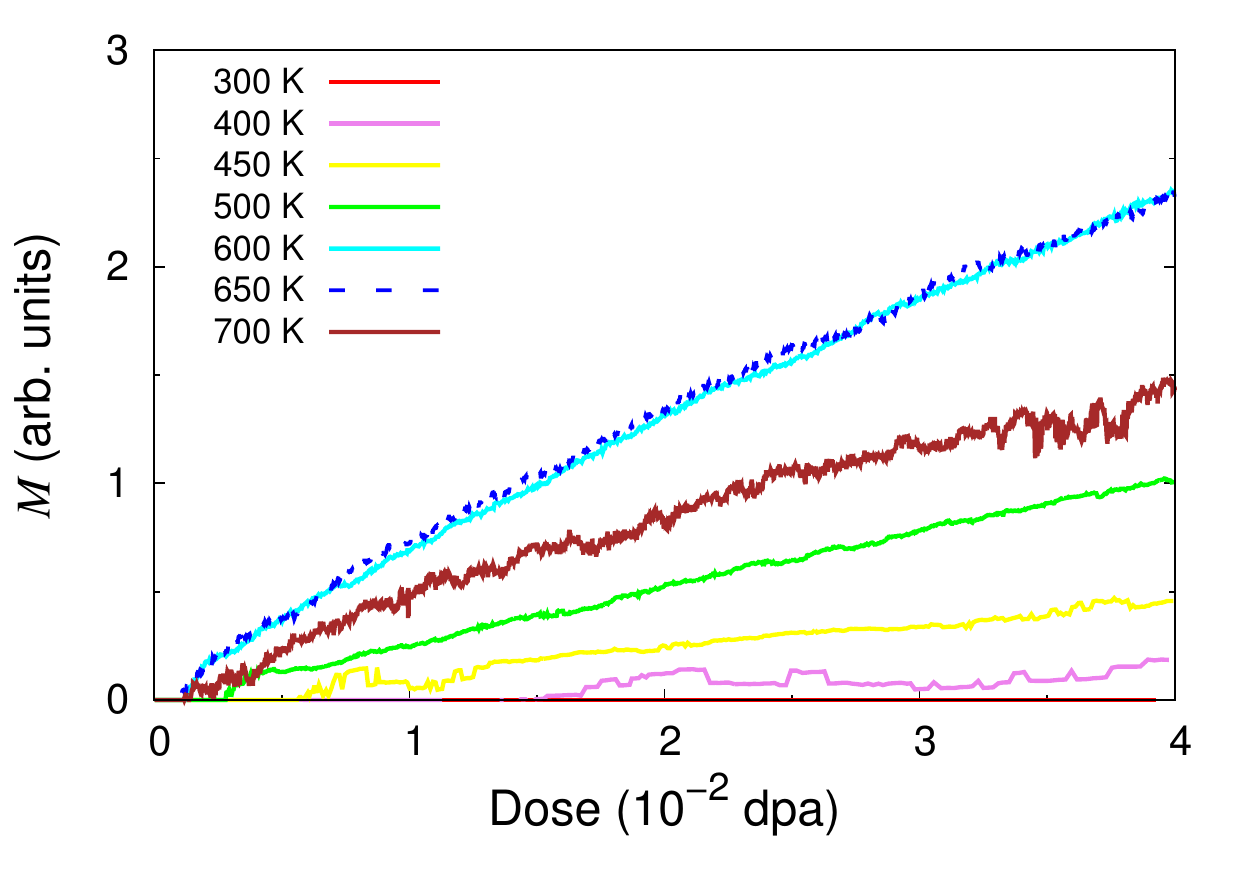}
	\caption{Evolution of the degree of segregation at different temperatures. The total solute concentration is 10\% at. The dose rate is 1.0 dpa$\cdot$s$^{-1}$.}
	\label{figDEGseg}
\end{figure}

Our kMC simulations are run up to a maximum dose of 0.04 dpa. The spatial solute concentration profiles along the $x$-dimension at 650 K as a function of dose are shown in Fig.\ \ref{figSRFprofile}. From the figure, the enrichment of solute atoms near the surfaces can be clearly appreciated, which is accompanied by local depletion in the subsurface region. Segregation near the surfaces increases with dose, in agreement with Dubey and El-Azab's work. These authors also studied the degree of segregation as a function of time $M(t)$, defined as:
\begin{equation}
	M(t)= \int_{0}^{l_{s}} \left(C(x, t)-\bar{C}\right)dx
\end{equation}
where $l_s$ is an arbitrary segregation distance, $C(x,t)$ is the instantaneous solute concentration profile, and $\bar{C}$ is the average solute concentration of the whole system. 
Here, we replace the integral by a discrete sum over lattice positions, with $l_s$ defined as the distance from the surface at which the local concentration is within 10\% of the background global concentration. To avoid noise due to lattice fluctuations, we apply a Savitzky-Golay smoothing filter \cite{1964ACSavitzky} prior to the determination of $l_s$.
The evolution of $M$ as a function of dose and temperature is shown in Fig.\ \ref{figDEGseg}. Our results are in agreement with those of Dubey and El-Azab, with $M$ increasing with dose monotonically in all cases. 
However, the evolution with temperature shows two distinct trends. First, $M$ increases with temperature up to a critical value of approximately 650 K. Then, it gradually decreases until, at $T=900$ K, the degree of segregation is practically zero.
The causes behind this behavior are well understood \cite{1979JoNMOkamoto}. Essentially, at low temperatures, vacancy mobility is limited, leading to high excess vacancy concentration and high recombination rates. As a consequence, segregation is low due to small defect fluxes to surfaces. At higher temperatures, vacancy and interstitial diffusion are activated resulting in net solute segregation. However, above 650 K, significant numbers of vacancies start to be emitted from the surfaces, leading to high back diffusion rates and again high recombination rates. The two effects result in a reduced solute segregation to the surfaces. Therefore, the maximum degree of segregation occurs at intermediate temperatures, consistent also with Dubey and El-Azab's findings.

\begin{figure}[h]
	\centering
	\begin{subfigure}{0.45\textwidth}
		\centering
		\includegraphics[width=\textwidth]{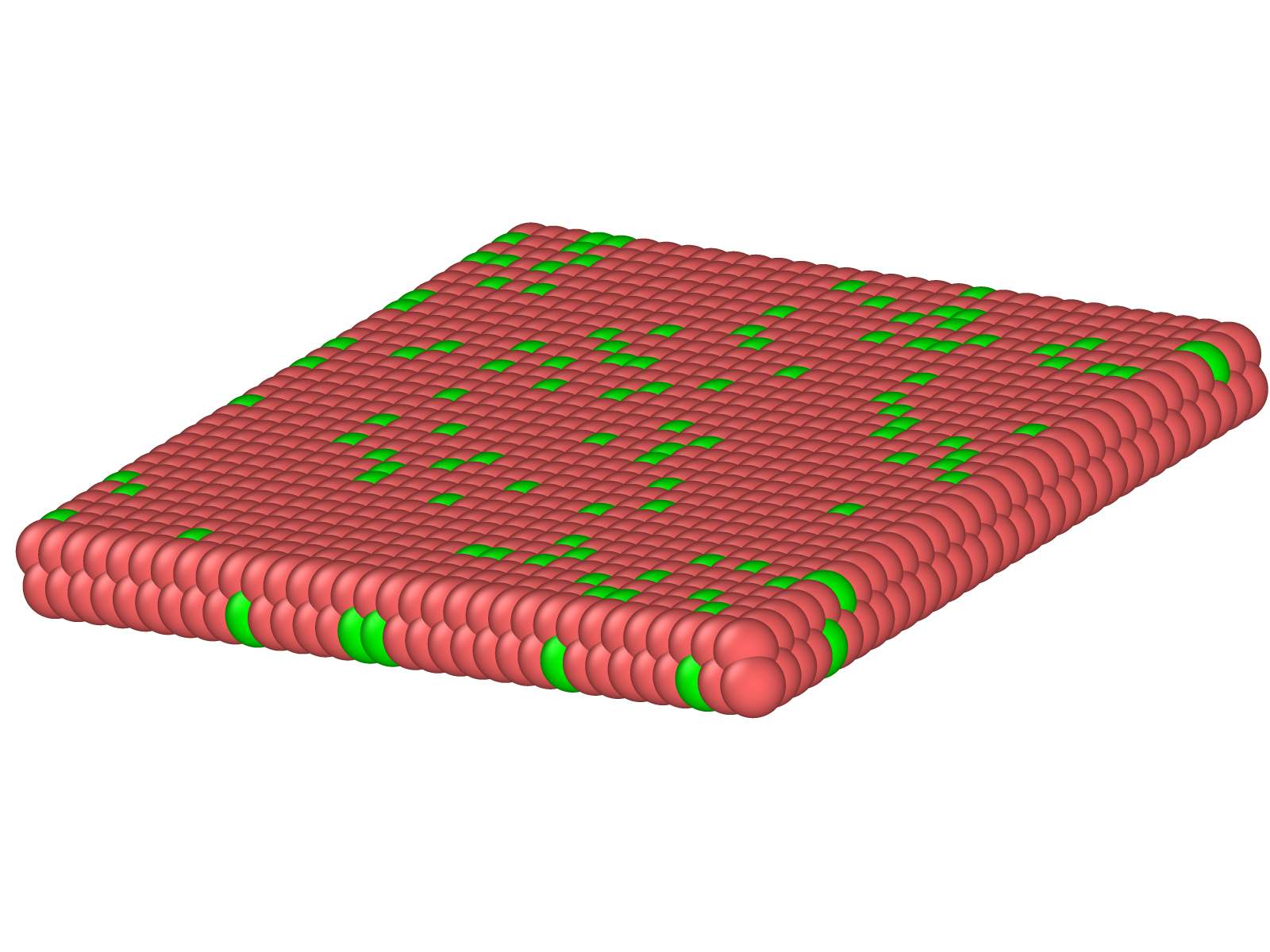}
		\caption{}
	\end{subfigure}
	\begin{subfigure}{0.45\textwidth}
		\centering
		\includegraphics[width=\textwidth]{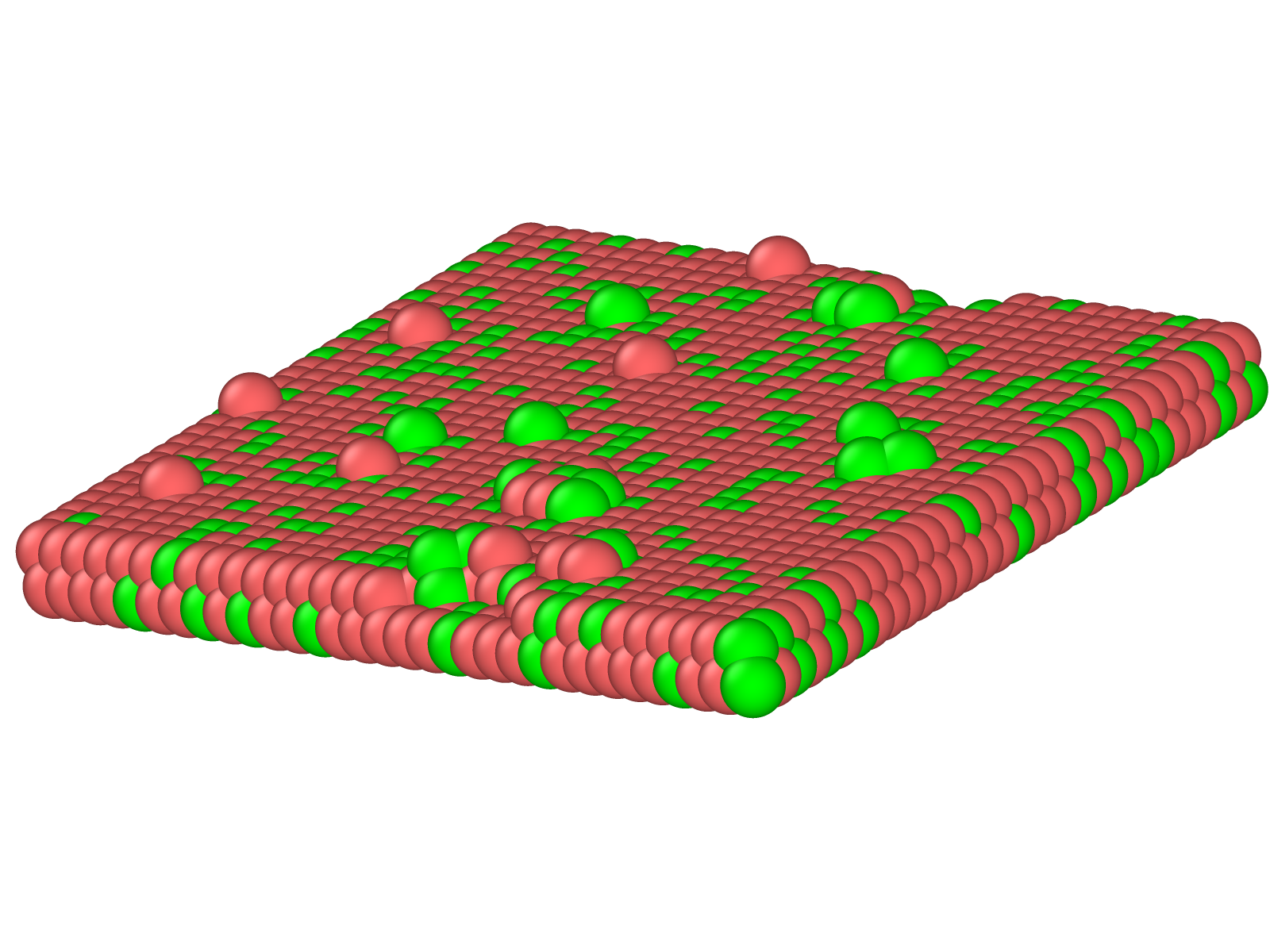}
		\caption{}
	\end{subfigure}
	\caption{Snapshots of surface roughness at (a) $t=0$, and (b) $t=0.02$ s for the Au-10.0 at.\ Cu system alloy at 500 K. Red dots represent solvent (A) atoms, while solute atoms (B) are represented as green dots. The dose rate is 1.0 dpa$\cdot$s$^{-1}$.}
	\label{fig_SRFsnap}
\end{figure}

KMC simulations are capable of providing morphological features that continuum methods cannot furnish. For example, our method can be used to study the evolution of the surface roughness, an example of which is shown in Fig. \ref{fig_SRFsnap}. The figure contains two snapshots of the surface at 500 K at different accumulated doses, where clear surface morphology changes can be appreciated. As well, surface roughness is accompanied by a concomitant increase in the concentration of solute atoms, which occurs by the mechanisms explained above.

\section{Summary and Conclusions}\label{concl}
We have proposed an extension of the standard ABV Hamiltonian to discrete binary systems containing interstitial defects. The chosen framework for this extension is the Ising model, where three new values for the spin variables are considered: `$+2$', representing pure self-interstitials (A-A), `$-2$', representing pure solute interstitials (B-B), and `$0$', for mixed interstitials (A-B). The reason for choosing these values is to preserve one of the essential magnitudes of the Ising model, the magnetization $N^{-1}\sum_i\sigma_i$, or, in the ABVI context, the excess solute concentration. The main advantage behind expressing a cluster expansion Hamiltonian as an Ising Hamiltonian is that thermodynamic information about the system can more easily be construed in the Ising framework. For example, the value of the constants of class 3 identified in eq.\ \eref{nonconf} uniquely determine the thermodynamic phase diagram of the ABVI model (much like constant $J$ in eq.\  \eref{habv} determines the structure of the ABV system). Indeed, one of the aspects of greatest interest associated with the ABVI model is to study how the presence of interstitials alters the behavior of substitutional binary alloys.

However, we leave this thermodynamic analysis for a specific binary system with well characterized bond energetics for a future study, and, instead, in this paper we have focused on verification by comparing against a number of selected published studies. The main tests that we have conducted include discrete lattice ABV and ABVI for dilute Fe-Cu alloys, as well as comparison against a spatially-resolved mean-field study of solute segregation at free surfaces in irradiated Au-Cu alloys. In all cases, basic metrics related to the timescale and/or some governing kinetic parameters were reproduced with good agreement. In terms of computational cost, our Ising ABVI model scales in a similar manner as second-order cluster expansion Hamiltonians with similar cutoff radius --as it should, given that no advantage is lost by simply recasting a cluster expansion Hamiltonian into the Ising form.

Thus, in conclusion, we present an ABVI Hamiltonian, cast as an Ising model Hamiltonian, for discrete event simulations that can be considered a generalization of ABV models. Our model has been verified against existing parameterizations of cluster expansion Hamiltonians using kinetic Monte Carlo simulations, with good agreement observed. We will study the thermodynamic behavior of our Hamiltonian in a future publication.
\section*{Acknowledgments}
We acknowledge support from DOE's Office of Fusion Energy Sciences via the Early Career Research Program.

\section*{References}

\end{document}